\documentclass[galaxies,article,accept,pdftex,moreauthors]{Definitions/mdpi} 
\usepackage{threeparttable}
\firstpage{1} 
\pubvolume{1}
\issuenum{1}
\articlenumber{0}
\pubyear{2026}
\copyrightyear{2026}
\datereceived{ } 
\daterevised{ } 
\dateaccepted{ } 
\datepublished{ } 

\nolinenumbers
\Title{Timing and Spectral Analysis of the 2024 Outburst of 2S 1553$-$542 with \textit{NuSTAR} and \textit{NICER}
}

\Author{Haifan  Zhu $^{1,}$*\orcidA{}, Wei Wang $^{2,}$*\orcidB{},Wen Yang $^{2}$, Mariano M\'endez$^{1}$, Chenxu Gao $^{3}$ ,Ziyi Xu $^{4}$, and Pengfu Tian $^{4}$}

\AuthorNames{Haifan Zhu, Wei Wang, Wen Yang, Mariano M\'endez, Chenxu Gao, Ziyi Xu,   and Pengfu Tian}

\address{%
$^{1}$ \quad Kapteyn Astronomical Institute, University of Groningen, P.O. BOX 800, 9700 AV Groningen, The Netherlands; zhu@astro.rug.nl\\
$^{2}$ \quad Department of Astronomy, School of Physics and Technology, Wuhan University, Wuhan 430072, China;\\
$^{3}$  \quad Shanghai Astronomical Observatory, Chinese Academy of Sciences, Nandan Road, Shanghai 200030, China;\\ 
$^{4}$  \quad Department of Physics, Guizhou Minzu University, Guiyang 550025, People's Republic of China
}

\corres{Correspondence: zhu@astro.rug.nl (H.Z.); wangwei2017@whu.edu.cn (W.W.)}

\abstract{
 We report a timing and spectral study of the 2024 outburst of the Be/X-ray binary pulsar 2S~1553$-$542 using \textit{NuSTAR} and \textit{NICER} observations. From the \textit{NuSTAR} light curve we measure a pulse period of $9.285022\pm0.000001$~s. The energy-resolved pulse profiles are dominated by a single peak and show a wing-like structure most clearly in the $12$--$22$~keV band. The pulsed fraction remains above 60\% and increases with energy. The phase-averaged \textit{NuSTAR} spectrum is described by an absorbed blackbody plus cutoff power-law continuum, together with an iron emission line and a cyclotron absorption feature. Using the \texttt{cyclabs} model, we obtain a cyclotron energy of $E_{\rm cyc}\simeq24.1$~keV, corresponding to a magnetic field strength of $B\sim3\times10^{12}$~G. Phase-resolved spectroscopy shows that the continuum and cyclotron-line parameters vary with pulse phase, and that the line becomes poorly constrained around the pulse-wing phase. We also searched the short \textit{NICER} GTIs for transient mHz variability using wavelet analysis and a CEEMDAN-based Hilbert--Huang transform. Localized excesses near $\sim10$~mHz and $\sim20$~mHz are found, but the short exposures, COI effects, red-noise fluctuations, and the lack of a well-constrained Fourier peak limit their significance. We therefore treat them as candidate mHz variability rather than firm mHz QPO detections.
}


\begin{document}


\section{Introduction}
\label{sec:intro}

X-ray binary pulsars, typically consisting of a highly magnetized neutron star accreting from a stellar companion, serve as premier laboratories for studying accretion physics under extreme conditions. A major subclass, Be/X-ray binaries (BeXBs), hosts an early-type Be star with a circumstellar decretion disk \citep{reig2011x,walter2015high}. These systems are characterized by two distinct outburst modes: periodic Type~I outbursts ($L_{\mathrm{X}} \sim 10^{36-37} \mathrm{erg s^{-1}}$) occurring near periastron passage, and rare, giant Type~II outbursts ($L_{\mathrm{X}} > 10^{37} \mathrm{erg s^{-1}}$) associated with major disk disruption events \citep{stella1986intermittent,okazaki2001natural}.

In the timing domain, mHz quasi-periodic oscillations (QPOs) are critical variability features that probe inner accretion flow dynamics and disk-magnetosphere interactions \citep{klis2000millisecond}. However, characterizing these features in X-ray pulsars presents significant challenges. Standard Fourier techniques, which assume signal stationarity, are often ill-suited to capture the transient and evolving nature of mHz QPOs \citep{yang2025millihertz}. Consequently, advanced time--frequency analysis techniques have become indispensable. Wavelet analysis, which offers a balance between time and frequency resolution, has proven to be a powerful tool for resolving transient signals in black hole systems \citep{chen2022wavelet,chen2022waveleta,jin2024wavelet,zhu2025timing}. Inspired by these successes, this method has recently been extended to neutron star systems. Notably, recent studies have successfully employed wavelet analysis to detect mHz QPO signals in X-ray pulsars such as Her~X-1 \citep{yang2025observations} and IGR~J19294+1816 \citep{yang2025detection}. These works demonstrate the unique advantage of wavelet techniques in uncovering transient low-frequency oscillatory features and complex sideband structures that are often inaccessible to global Fourier analysis. 

Although wavelet analysis represents a significant advance, it is subject to the cone of influence (COI), which can limit the reliability of low-frequency detection in observations with short exposure times. To address this, the Hilbert--Huang Transform (HHT) has emerged as a robust alternative for characterizing nonlinear and non-stationary processes \citep{huang1998empirical,yu2023hilbert,shui2024phase,zhu2026timing}. By utilizing the Empirical Mode Decomposition (EMD) and its variants, HHT adapts to the local characteristic time scale of the data, eliminating the windowing limitations inherent to wavelet and Fourier methods. A pioneering application in this field was demonstrated by \citet{hsieh2020phase} in their study of 4U~1636--53. By utilizing HHT to extract the instantaneous phase of intermittent mHz QPOs, they successfully constructed phase-resolved spectra for non-stationary signals---a task challenging for standard period-folding techniques. Such advanced timing analyses are crucial for distinguishing between geometric effects and intrinsic luminosity variations in the accretion flow.

Spectrally, a cyclotron resonant scattering feature (CRSF) allows for a direct estimation of the neutron star's magnetic field via resonant scattering in quantized Landau levels \citep{trumper1978evidence}. The surface magnetic field is related to the line energy $E_{\mathrm{cyc}}$ by $B \simeq E_{\mathrm{cyc}}(1+z)/11.57 \times 10^{12} \mathrm{G}$ \citep{meszaros1992high}. Such features have been widely detected in accreting pulsars, providing key diagnostics of the accretion column geometry \citep{santangelo1998bepposax,liu2022variations}.

The transient BeXB 2S~1553$-$542, discovered in 1975 \citep{walter1976mx1553}, has an orbital period of $\sim30.6$~d and a pulse  period of $\sim9.3$~s \citep{kelley19822s}. Historical observations have been crucial for understanding this source. The 2015 giant outburst revealed a CRSF at $\sim23.5$~keV, implying $B \sim 3\times10^{12}$~G \citep{tsygankov2016nustar,lutovinov20162s}. Subsequently, the 2021 outburst showed a luminosity-dependent CRSF energy, suggesting a transition near the critical luminosity \citep{malacaria2022accreting}.

On 2024 September 6, \textit{MAXI} detected a new outburst from 2S~1553$-$542 \citep{2024ATel16835....1N}, which was followed by \textit{NuSTAR} and \textit{NICER} observations. In this paper, we present a timing and spectral analysis of this
outburst. Section~\ref{obs} describes the observations and data
reduction. Section~\ref{RESULTS} presents the timing and spectral
results. Sections~\ref{DISCUSSION} and \ref{conclusion} provide the
discussion and conclusions, respectively.
\section{Observations and Data Reduction}
\label{obs}
\subsection{Observations}
\textit{NuSTAR} \citep{harrison2013nuclear}, launched on June 13, 2012, is the first focusing high-energy X-ray telescope in orbit, operating in the 3 to 79 keV range. \textit{NuSTAR}'s primary scientific objectives include investigating obscured active galactic nuclei, studying compact objects in the Milky Way, examining non-thermal radiation in supernova remnants, observing blazars in coordination with other telescopes, and analyzing emissions from supernovae to constrain explosion models. On October 7, 2024, \textit{NuSTAR} conducted an observation of 2S~1553$-$542. The \textit{NuSTAR} observation is listed in the first row of
Table~\ref{2STAB}.

\textit{NICER} \citep{gendreau2016neutron}, a mission of opportunity aboard the International Space Station (ISS), was selected under NASA’s Explorers Program as the agency’s first dedicated mission for neutron star research. Installed on the ISS on June 13, 2017, following its June 3 launch, \textit{NICER} began operations nearly 50 years after neutron stars were first discovered. As an X-ray telescope, it is equipped with silicon-drift detectors that cover an energy range of 0.2 to 12 keV, providing an effective collecting area exceeding 2000 cm² at 1.5 keV and 600 $\rm cm^2$ at 6 keV. By capturing individual X-ray photons with a time-tagging resolution of 300 nanoseconds, \textit{NICER} enables high-precision measurements of neutron star energetics, structure, and dynamics, facilitating the study of rapid brightness and spectral fluctuations in the soft X-ray band. In 2024, NICER conducted five observations of 2S 1553-542. Detailed information on these observations is listed in Table~\ref{2STAB}.

\begin{table*}
    \centering
    \caption{The observation information of 2S 1553-542.}
    \label{2STAB}
    \renewcommand{\arraystretch}{1.3} 

    \begin{threeparttable} 
        \begin{tabular}{ccccc}
            \toprule \toprule
            Instrument & ObsID & Start Time & Exposure & Pulse Period \\
             & & (UTC) & (s) & (s) \\
            \midrule 
            \textit{NuSTAR} & 91001343002 & 2024-10-07 12:06:12.00 & 33565 & $9.285022\pm0.000001$ \\
            \textit{NICER}  & 7202030101  & 2024-10-03 01:23:00.00 & 1012  & $9.2849\pm0.00087$    \\
            \textit{NICER}  & 7202030102  & 2024-10-05 03:03:30.00 & 1243  & $9.28429\pm0.00166$   \\
            \textit{NICER}  & 7202030103  & 2024-10-07 19:49:18.00 & 268   & $9.28325\pm0.00276$   \\
            \textit{NICER}  & 7202030104  & 2024-10-08 02:01:39.00 & 342   & $9.28338\pm0.00051$   \\
            \textit{NICER}  & 7202030105\tnote{a} & 2024-11-20 01:52:08.00 & 0 & \textemdash \\
            \bottomrule 
        \end{tabular}
        
        \begin{tablenotes}
            \footnotesize
            \item[a] ObsID~7202030105 was excluded due to insufficient exposure time.
        \end{tablenotes}
    \end{threeparttable}
\end{table*}

\subsection{Data Reduction}

According to the standard procedures\footnote{https://heasarc.gsfc.nasa.gov/docs/nustar/analysis/}, we processed the \textit{NuSTAR} data using HEAsoft v6.32 with the calibration files v20251215. To extract the source spectra, a circular region with a radius of $120''$ centered on the source position was selected, and the background spectra were extracted from a source-free circular
region of the same size on the same detector as the source region. Standard data-processing tools were used to generate level-2
(\textit{nupipeline}) and level-3 (\textit{nuproducts}) products.
The \textit{NuSTAR} light curves were corrected for both the spacecraft
clock offset and instrumental dead time. The barycentric correction was
performed with the FTOOL \textit{barycorr}, using the \textit{NuSTAR}
orbit file and the CALDB clock-correction file
(\texttt{clockfile=CALDB}). The source and background light curves were
generated with \textit{nuproducts} using \texttt{correctlc=yes}, which
invokes \textit{nulccorr} and applies the instrumental livetime/dead-time
correction. The background contribution was subsequently subtracted from
the FPMA and FPMB light curves using the FTOOL \textit{lcmath}. 
To further investigate the energy-dependent characteristics of this source, we divided the \textit{NuSTAR} data into the following energy bands: 3-5 keV, 5-8 keV, 8-12 keV, 12-18 keV, 18-22 keV, 22-34 keV, 34-50 keV and 50-70 keV. The unequal energy intervals were chosen to balance energy
resolution and signal-to-noise ratio: narrower bands were used at
lower energies and around the CRSF, while wider bands were required
at higher energies because of the decreasing count rate.

For the NICER data analysis, we followed the official analysis threads\footnote{https://heasarc.gsfc.nasa.gov/docs/nicer/analysis\_threads}, utilizing HEAsoft V6.32 software and the calibration files (xti20221001). The tools \textit{nicer-l2} and \textit{nibackgen3C50} were used to extract the source and background, respectively. Additionally, light curves in the 1-10 keV energy range were generated using the \textit{nicerl3-lc} tool. 
\section{ANALYSIS AND RESULTS}
\label{RESULTS}
\subsection{Pulse Profile and Pulsed Fraction }
To obtain the pulse profile, we processed the light curves accordingly. The \textit{NuSTAR} and \textit{NICER} light curves were barycenter corrected using the \texttt{barycorr} tool, with a time bin size of 0.0078125 s. The pulse periods were derived from the combined, background-corrected \textit{NuSTAR} FPMA and FPMB light curves. 

The epoch-folding technique \citep{leahy1983searches} was applied using the \texttt{efsearch} tool\footnote{https://heasarc.gsfc.nasa.gov/xanadu/xronos/help/efsearch.html} from HEASARC to determine the pulse period. Additionally, we employed \texttt{stingray}\footnote{\url{https://github.com/StingraySoftware/stingray}} \citep{bachetti2024stingray} to fit the stable pulse frequency by modeling the pulse peak with Gaussian functions.
This gives a pulse period of $9.285022\pm0.000001$~s for the \textit{NuSTAR} observation.  However, for the \textit{NICER} observations, we found that the detected pulse peak exhibited variability, likely due to the limited observation duration and potential short-term variations in the emission. This variability introduced challenges in determining a stable pulse period. To account for this, we performed individual fits for each \textit{NICER} observation, and we present the best-fit results in the last column of Table~\ref{2STAB}. These variations highlight the necessity of longer or more continuous observations to better characterize the pulsation behavior in future studies. 

\begin{figure*}
	\includegraphics[width=1\columnwidth]{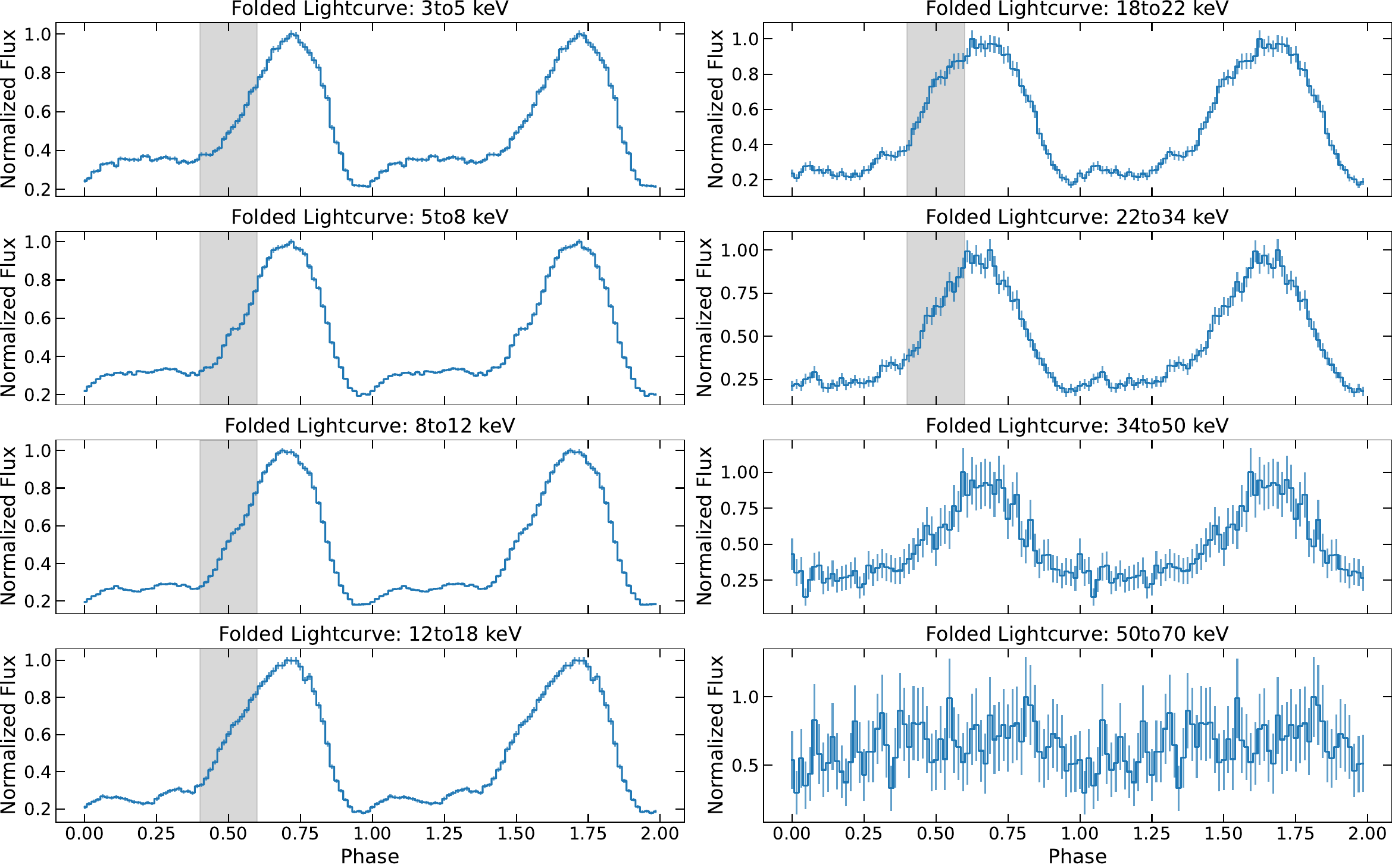}
    \caption{Pulse profiles from the NuSTAR observation, with each panel corresponding to a selected energy band indicated in the figure. The gray-shaded regions denote the locations of the wing structures.}
    \label{fig1}
\end{figure*}
\begin{figure}
	\includegraphics[width=0.7\textwidth]{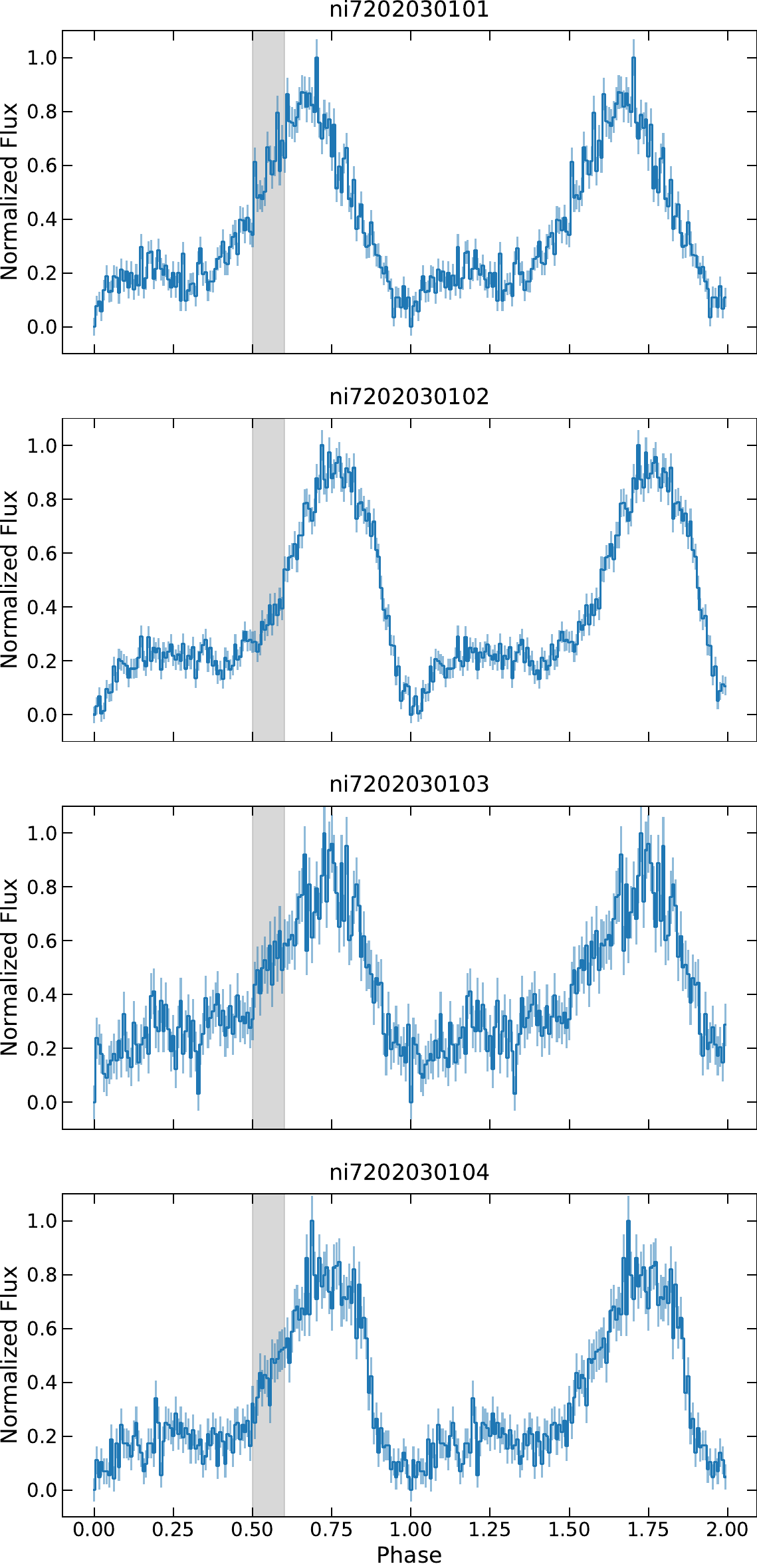}
    \caption{Pulse profiles from NICER (1-10 keV) , with each panel representing one observation. The corresponding observation IDs are indicated in the figure. The gray-shaded regions mark the locations of the wing structures.}
    \label{fig2}
\end{figure}

Using the derived pulse periods, we performed phase folding on the light curves from each observation to obtain the pulse profiles. The characteristics of the pulse profiles are strongly correlated with the geometric configuration of the neutron star’s surface emission regions-such as the distribution of hot spots and the inclination of the magnetic poles. Meanwhile, their energy-dependent behavior reflects the dominant role of Compton scattering processes occurring in the boundary layer of the accretion column.

To improve the statistical significance of the data, we did not analyze the light curves from NuSTAR FPMA and FPMB separately; instead, we combined them. The resulting pulse profiles are shown in Fig.~\ref{fig1}. The profiles exhibit a single-peaked structure with only weak dependence on energy. Fig.~\ref{fig1} presents the evolution of the pulse profiles in eight different energy bands (3 to 5, 5 to 8, 8 to 12, 12 to 18, 18 to 22, 22 to 34, 34 to 50, and 50 to 70 keV, as labeled in the figure).

Three main features can be identified: a primary maximum, a primary minimum, and an energy-dependent wing-like structure is observed between pulse
phases 0.4 and 0.6. The feature is not clearly visible in the
3--5~keV band, becomes weakly discernible at 5--8~keV, and is most
prominent in the 12--18 and 18--22~keV bands. It weakens in the
22--34~keV band and is no longer evident above 34~keV. Therefore,
the 12--22~keV range represents the energy interval in which the
wing-like structure is most pronounced. The obtained results are similar to those found by  \cite{tsygankov2016nustar} and \cite{malacaria2022accreting}. It should be noted that phase zero is defined at the pulse minimum.

The pulse profiles obtained from the four \textit{NICER}
observations (Fig.~\ref{fig2}) remain broadly single-peaked but show
modest evolution. Compared with the observation on October 3
(ObsID: 7202030101), the pulse maxima in the subsequent observations
occur at slightly later phases, and the shape of the descending
flank also changes. Although the normalized flux around phases
0.4--0.6 is comparable in some observations, no distinct wing-like
structure is evident in the first two observations. This feature
becomes apparent on October 7 (ObsID: 7202030103) and remains visible
on October 8. The October 7 observation was obtained close in time
to the \textit{NuSTAR} observation, and the feature appears at
approximately the same phase as the wing-like structure seen by
\textit{NuSTAR}, supporting an intrinsic origin. Given the short
exposures and the fact that each observation was folded separately,
we do not attach quantitative significance to the small shift of
the pulse maximum.

\begin{figure}
    \includegraphics[width=0.8\columnwidth]{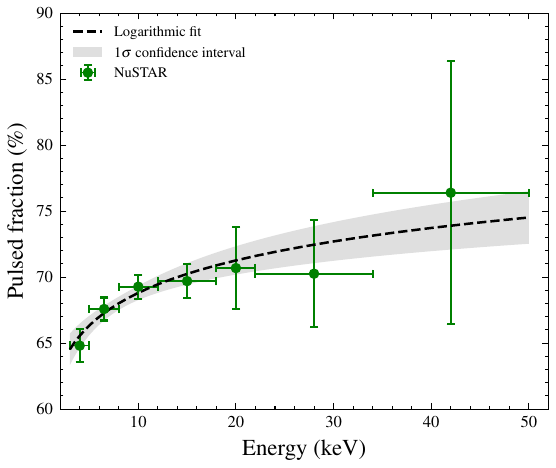}
    \caption{Energy dependence of the pulsed fraction derived from the
\textit{NuSTAR} data. The dashed line shows the best-fitting
logarithmic relation, with the gray shaded region indicating its
$1\sigma$ confidence interval.}

    \label{pf}
\end{figure}

Using NuSTAR observations, we further quantified the energy dependence of the pulsed fraction. The pulsed fraction (PF) is a key parameter characterizing the modulation strength of a periodic signal, defined as the ratio of the modulated component's amplitude to the total emission intensity. For X-ray pulsars, it can be expressed as the ratio of the difference between the maximum and minimum intensities in the pulse profile to their sum, namely:
    \begin{equation}
    \mathrm{PF} = \frac{I_{\max} - I_{\min}}{I_{\max} + I_{\min}}.
    \end{equation}
Fig.~\ref{pf} shows how the PF varies as a function of energy. The pulsed fraction remains at a consistently high level (above 60\%) and shows a steadily increasing trend throughout the energy range.  To quantify the energy dependence, we fitted the PF measurements
with the empirical logarithmic relation
\[
{\rm PF}(E)=a+b\log_{10}(E/10~{\rm keV}).
\]
The best-fitting parameters are $a=68.81\pm0.54\%$ and
$b=8.17\pm2.46$ percentage points per decade in energy. The
positive slope is significant at the $3.33\sigma$ level
($p=8.81\times10^{-4}$), demonstrating a statistically significant
overall increase in PF with energy. 
    
\subsection{Spectral Analysis}
\label{sec:spec_analysis}

To characterize the overall spectral properties, we analyzed the
time-averaged \textit{NuSTAR} spectrum over the 3--79~keV energy
range. We first fitted the spectrum using the T\"ubingen--Boulder
interstellar absorption model (\texttt{TBabs}), a blackbody
component (\texttt{bbodyrad}), an exponentially cutoff power law
(\texttt{cutoffpl}), and a Gaussian emission line (\texttt{gauss})
representing the iron K$\alpha$ complex. The baseline model
(Model~1) is defined as
\begin{equation}
\mathcal{M}_{\rm cont}
=
\texttt{constant}\times\texttt{TBabs}
\times
(\texttt{bbodyrad}+\texttt{cutoffpl}+\texttt{gauss}).
\end{equation}
Here, $\mathcal{M}_{\rm cont}$ denotes the complete baseline spectral
model. The multiplicative \texttt{constant} accounts for the
cross-calibration difference between FPMA and FPMB; it was fixed at
unity for FPMA and allowed to vary for FPMB. The \texttt{TBabs}
component describes interstellar photoelectric absorption. Its
equivalent hydrogen column density was fixed at
$N_{\rm H}=2.3\times10^{22}~{\rm cm}^{-2}$, following previous
studies \citep{lutovinov20162s}. The \texttt{component bbodyrad} is 
characterized by the temperature $kT$ and normalization
$K_{\rm bb}=(R_{\rm km}/D_{10})^{2}$, where $R_{\rm km}$ is the
radius of the emitting region in kilometers and $D_{10}$ is the
source distance in units of 10~kpc. The \texttt{cutoffpl} component
is characterized by the photon index $\Gamma$, cutoff energy
$E_{\rm cut}$, and normalization, while the \texttt{gauss}
component is described by its centroid energy $E_{\rm Fe}$, width
$\sigma_{\rm Fe}$, and normalization.

\begin{figure}
  \centering
      \includegraphics[width=0.6\columnwidth]{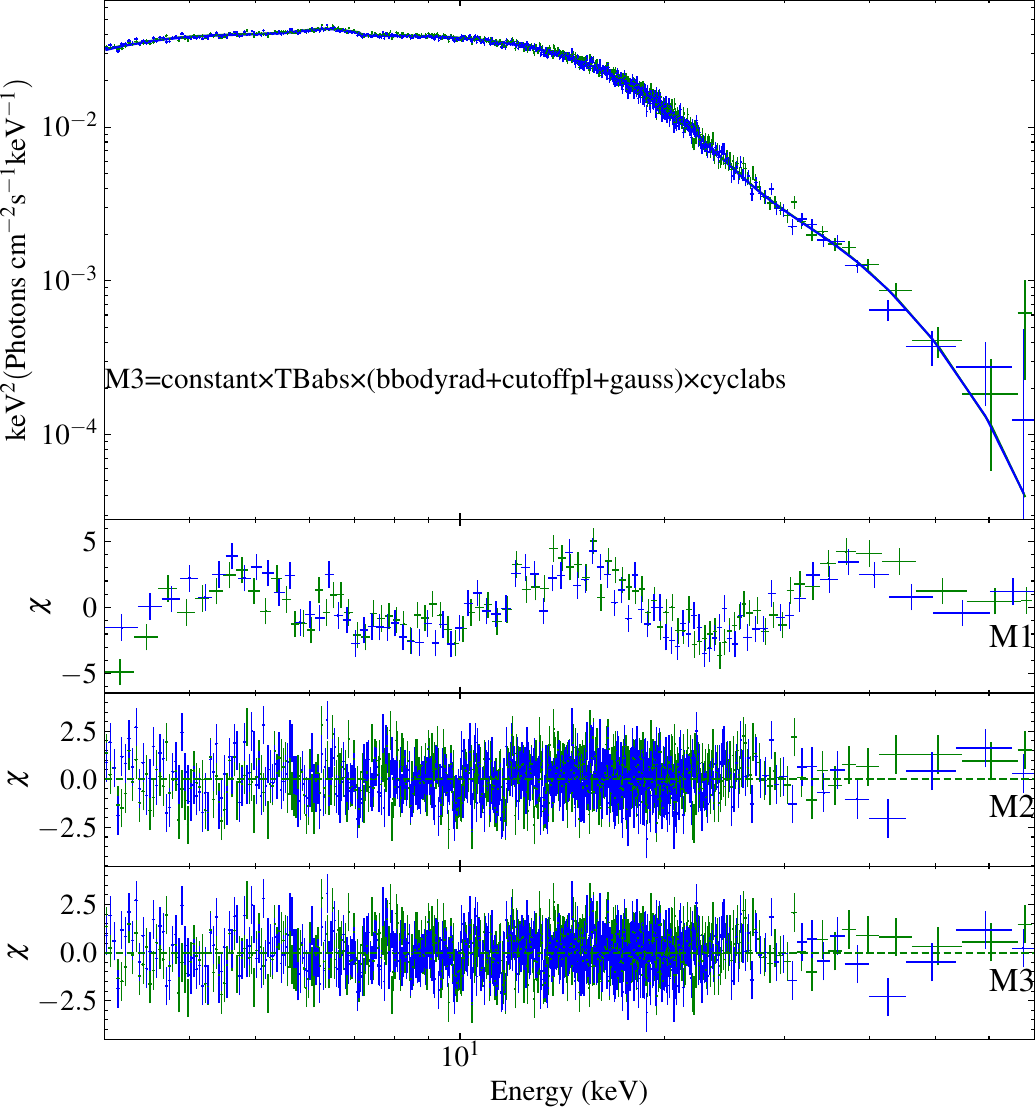}
      \caption{The phase-averaged \textit{NuSTAR} spectrum fitted with
Model~3 (\texttt{cyclabs}; upper panel), together with the residuals
for Model~1, Model~2 (\texttt{gabs}), and Model~3
(\texttt{cyclabs}) in the three lower panels. The FPMA and FPMB data
are shown in green and blue, respectively. Both CRSF models remove
the broad residual structure around 20--30~keV present in Model~1. The residuals have been rebinned for display purposes only.}
      \label{2sspec3}
\end{figure}

Model~1 yields $\chi^{2}/{\rm d.o.f.}=1898.16/1439$ and leaves
prominent absorption-like residuals around 20--30~keV
(Fig.~\ref{2sspec3}). To characterize the CRSF, we tested two
phenomenological absorption-line profiles. The \texttt{gabs}
model\footnote{\url{https://heasarc.gsfc.nasa.gov/docs/software/xspec/manual/node260.html}}
describes a Gaussian absorption profile in optical depth and is
characterized by the centroid energy $E_{\rm gabs}$, width
$\sigma_{\rm gabs}$, and line strength $S_{\rm gabs}$. The
\texttt{cyclabs}
model\footnote{\url{https://heasarc.gsfc.nasa.gov/docs/software/xspec/manual/node254.html}}
instead adopts a pseudo-Lorentzian cyclotron absorption profile,
parameterized by the line energy $E_{\rm cyc}$, width $W_{\rm cyc}$,
and depth $D_{\rm cyc}$. The resulting models and fits are:

\begin{itemize}
    \item Model~2:
    $\mathcal{M}_{\rm cont}\times\texttt{gabs}$, which gives
    $\chi^{2}/{\rm d.o.f.}=1388.11/1436$ and
    $E_{\rm gabs}=27.95\pm0.34$~keV.

    \item Model~3:
    $\mathcal{M}_{\rm cont}\times\texttt{cyclabs}$, which gives
    $\chi^{2}/{\rm d.o.f.}=1383.44/1436$ and
    $E_{\rm cyc}=24.11\pm0.23$~keV.
\end{itemize}

Relative to Model~1, the addition of \texttt{gabs} and
\texttt{cyclabs} improves the fit by $\Delta\chi^{2}=510.05$ and
$514.72$, respectively, for three additional free parameters.
Both profiles therefore provide a substantial improvement over the
baseline model. The difference between the fitted line energies
arises from the different functional forms of \texttt{gabs} and
\texttt{cyclabs}.

Using the \texttt{cyclabs} result, we estimate a magnetic field
strength in the line-forming region of
$B\approx3\times10^{12}$~G, assuming $z\approx0.3$. The unabsorbed 3--70~keV flux is
approximately
$1.06\times10^{-9}~{\rm erg~s^{-1}~cm^{-2}}$. For the reported
distance range of 16--24~kpc, this corresponds to
$L_{\rm X}\approx(3.3-7.3)\times10^{37}~{\rm erg~s^{-1}}$.
The best-fitting parameters are listed in
Table~\ref{2SPARA}.

\begin{table*}[ht]
\centering
\caption{Best-fitting parameters of the phase-averaged
\textit{NuSTAR} spectrum of 2S~1553$-$542 obtained with the three
spectral models. The quoted uncertainties correspond to the
$1\sigma$ statistical errors.}
\label{2SPARA}

\renewcommand{\arraystretch}{1.05}
\setlength{\tabcolsep}{1.5pt}

\begin{tabularx}{\textwidth}{ccccc}
\hline
Parameter & Unit & Model 1 & Model 2 & Model 3 \\
\hline

$C_{\rm FPMB}$
& --
& $0.991\pm0.001$
& $0.992\pm0.001$
& $0.992\pm0.001$ \\

$N_{\rm H}$
& $10^{22}~{\rm cm}^{-2}$
& $2.3$ (fixed)
& $2.3$ (fixed)
& $2.3$ (fixed) \\

$kT$
& keV
& $3.260\pm0.009$
& $0.861\pm0.017$
& $0.876\pm0.016$ \\

$K_{\rm bb}$
& --
& $0.607\pm0.010$
& $26.1\pm1.6$
& $24.7\pm1.5$ \\

$\Gamma$
& --
& $1.946\pm0.060$
& $-0.564\pm0.048$
& $-0.503\pm0.067$ \\

$E_{\rm cut}$
& keV
& $20.8\pm4.3$
& $5.371\pm0.105$
& $5.868\pm0.215$ \\

$K_{\rm cutoffpl}$
& ${\rm photons~keV^{-1}~cm^{-2}~s^{-1}}$
& $0.1255\pm0.0070$
& $0.0069\pm0.0005$
& $0.0070\pm0.0005$ \\

$E_{\rm Fe}$
& keV
& $6.200\pm0.089$
& $6.298\pm0.003$
& $6.304\pm0.004$ \\

$\sigma_{\rm Fe}$
& keV
& $0.489\pm0.057$
& $0.456\pm0.048$
& $0.472\pm0.048$ \\

$K_{\rm Fe}$
& $10^{-4}~{\rm photons~cm^{-2}~s^{-1}}$
& $5.28\pm0.63$
& $5.78\pm0.64$
& $5.82\pm0.65$ \\

$E_{\rm gabs}$
& keV
& --
& $27.95\pm0.34$
& -- \\

$\sigma_{\rm gabs}$
& keV
& --
& $6.22\pm0.32$
& -- \\

$S_{\rm gabs}$
& keV
& --
& $10.3\pm1.1$
& -- \\

$E_{\rm cyc}$
& keV
& --
& --
& $24.11\pm0.23$ \\

$W_{\rm cyc}$
& keV
& --
& --
& $10.1\pm0.8$ \\

$D_{\rm cyc}$
& --
& --
& --
& $0.754\pm0.057$ \\

$F_{3-70}$
& $10^{-9}~{\rm erg~s^{-1}~cm^{-2}}$
& --
& $1.064\pm0.001$
& $1.064\pm0.001$ \\

$\chi^{2}/{\rm d.o.f.}$
& --
& $1898.16/1439$
& $1388.11/1436$
& $1383.44/1436$ \\

$\chi_{\nu}^{2}$
& --
& $1.319$
& $0.967$
& $0.963$ \\

\hline
\end{tabularx}

\begin{flushleft}
\footnotesize
The FPMA cross-normalization constant ($C_{FPMA}$)  was fixed at unity, while
$C_{\rm FPMB}$ was allowed to vary. Model~1 is the baseline model;
Models~2 and 3 additionally include \texttt{gabs} and
\texttt{cyclabs}, respectively.
\end{flushleft}

\end{table*}

To explore the geometric variability of the emission region, we performed a coarse phase-resolved spectral analysis. We divided the pulse period into 5 phase bins to ensure sufficient signal-to-noise ratio for the CRSF detection.We fitted the phase-resolved spectra using the phenomenological
\texttt{gabs} profile (Model~2) to investigate the phase-dependent
evolution of the cyclotron-line parameters. This choice was
motivated by two considerations. First, no secondary cyclotron
absorption feature associated with the second-harmonic term of the
full \texttt{cyclabs} formulation was evident in the phase-averaged
spectrum. Second, the cyclotron feature had a relatively low
signal-to-noise ratio in some phase bins, making any additional
harmonic parameters difficult to constrain. We therefore adopted a
single \texttt{gabs} component for the phase-resolved analysis. This
choice also facilitates a direct comparison with the previous
phase-resolved study of the 2021 outburst of
2S~1553$-$542, in which the CRSF was described using the same line
profile \citep{rai2023spectral}. The phase-dependent line parameters
reported below therefore refer to the \texttt{gabs}
parameterization.

 \begin{figure}
    \centering

        \includegraphics[width=0.9\columnwidth]{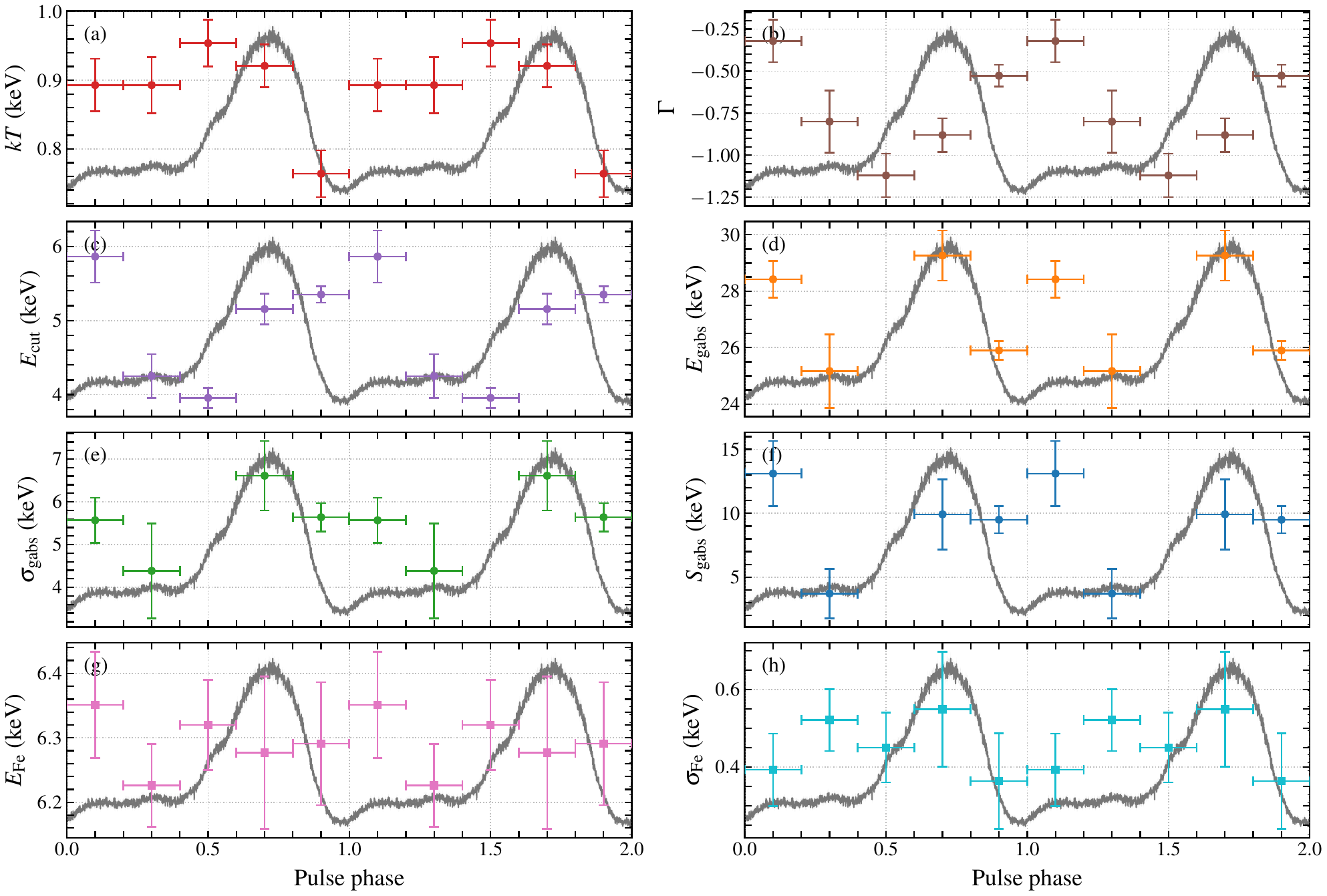}
       \caption{Phase-resolved variations of the \textit{NuSTAR} spectral
parameters, shown over two pulse cycles for clarity. From left to
right and top to bottom, the panels show the blackbody temperature
$kT$, photon index $\Gamma$, cutoff energy $E_{\rm cut}$,
\texttt{gabs} centroid energy $E_{\rm gabs}$, line width
$\sigma_{\rm gabs}$, line strength $S_{\rm gabs}$, iron-line
centroid energy $E_{\rm Fe}$, and iron-line width
$\sigma_{\rm Fe}$. The colored data points show the best-fitting
spectral parameter values; the vertical error bars indicate their
$1\sigma$ uncertainties, while the horizontal error bars show the
phase-bin widths. The gray curve in each panel shows the normalized
\textit{NuSTAR} pulse profile.}

        \label{2scy}

    \end{figure}

We observed significant modulation of the spectral parameters across the pulse cycle as shown in Fig.~\ref{2scy}. Notably, the non-thermal components ($\Gamma$ and $E_{\rm cut}$) show an anti-correlation with the pulse intensity, while the CRSF parameters (centroid energy and width) peak near the pulse maximum. In the phase intervals 0.4--0.6, no significant CRSF is detected using the adopted spectral model, and therefore its
parameters cannot be constrained. Consequently, no data
points are shown for this phase interval in panels (d)--(f) of
Fig.~\ref{2scy}. This strong phase dependence indicates a complex emission geometry and magnetic field configuration during the neutron star's rotation \citep{staubert2019cyclotron}, consistent with the theoretical expectations for a supercritical accretion column. However, the fundamental magnetic field estimate derived from the phase-averaged spectrum remains robust for the purpose of our timing analysis.

\begin{table*}[ht]
\centering
\caption{Best-fitting parameters from the phase-resolved
\textit{NuSTAR} spectral analysis using Model~2
($\mathcal{M}_{\rm cont}\times\texttt{gabs}$).}
\label{tab:phase_gabs}

\renewcommand{\arraystretch}{1.5}
\setlength{\tabcolsep}{0.80pt}

\begin{tabular}{lccccc}
\hline
Parameter
& phase 1
& phase 2
& phase 3
& phase 4
& phase 5 \\
\hline

$C_{\rm FPMB}$
& $0.995\pm0.005$
& $0.993\pm0.006$
& $0.992\pm0.005$
& $0.981\pm0.004$
& $0.994\pm0.003$ \\

$N_{\rm H}$ ($10^{22}~{\rm cm}^{-2}$)
& $2.3$ (fixed)
& $2.3$ (fixed)
& $2.3$ (fixed)
& $2.3$ (fixed)
& $2.3$ (fixed) \\

$kT$ (keV)
& $0.892\pm0.039$
& $0.893\pm0.041$
& $0.954\pm0.037$
& $0.898\pm0.034$
& $0.764\pm0.034$ \\

$K_{\rm bb}$
& $17.8\pm2.4$
& $18.4\pm2.7$
& $16.8\pm1.8$
& $28.5\pm3.5$
& $61.9\pm9.4$ \\

$\Gamma$
& $-0.317\pm0.125$
& $-0.810\pm0.184$
& $-1.121\pm0.129$
& $-0.907\pm0.107$
& $-0.527\pm0.065$ \\

$E_{\rm cut}$ (keV)
& $5.88\pm0.36$
& $4.36\pm0.30$
& $3.95\pm0.13$
& $5.13\pm0.22$
& $5.35\pm0.12$ \\

$K_{\rm cutoffpl}$ ($10^{-3}$)
& $7.21\pm1.46$
& $3.27\pm0.94$
& $2.20\pm0.48$
& $3.93\pm0.70$
& $15.49\pm1.75$ \\

$E_{\rm Fe}$ (keV)
& $6.35\pm0.08$
& $6.23\pm0.06$
& $6.32\pm0.07$
& $6.28\pm0.12$
& $6.29\pm0.10$ \\

$\sigma_{\rm Fe}$ (keV)
& $0.39\pm0.09$
& $0.52\pm0.08$
& $0.45\pm0.09$
& $0.55\pm0.15$
& $0.36\pm0.12$ \\

$K_{\rm Fe}$ ($10^{-4}$)
& $4.57\pm1.04$
& $7.76\pm1.39$
& $5.36\pm1.17$
& $6.14\pm2.06$
& $5.26\pm1.67$ \\

$E_{\rm gabs}$ (keV)
& $28.44\pm0.64$
& $25.16\pm1.31$
& $--$
& $29.30\pm0.94$
& $25.90\pm0.33$ \\

$\sigma_{\rm gabs}$ (keV)
& $5.58\pm0.52$
& $4.36\pm1.12$
& $--$
& $6.75\pm0.89$
& $5.64\pm0.33$ \\

$S_{\rm gabs}$ (keV)
& $13.18\pm2.55$
& $3.69\pm1.96$
& $--$
& $10.21\pm3.02$
& $9.50\pm1.06$ \\

$\chi^{2}/{\rm d.o.f.}$
& $880.06/942$
& $867.36/881$
& $968.89/923$
& $1065.00/1093$
& $1072.08/1139$ \\

$\chi^{2}_{\nu}$
& $0.934$
& $0.985$
& $1.05$
& $0.974$
& $0.941$ \\

\hline
\end{tabular}

\begin{flushleft}
\footnotesize
The FPMA cross-normalization constant was fixed at unity, while
$C_{\rm FPMB}$ was allowed to vary. The hydrogen column density was
fixed at $N_{\rm H}=2.3\times10^{22}~{\rm cm}^{-2}$.
$K_{\rm cutoffpl}$ is given in units of
$10^{-3}~{\rm photons~keV^{-1}~cm^{-2}~s^{-1}}$, and
$K_{\rm Fe}$ in units of
$10^{-4}~{\rm photons~cm^{-2}~s^{-1}}$.
The quoted uncertainties correspond to the $1\sigma$ statistical
errors. Parameters marked with $\dagger$ are poorly constrained.
\end{flushleft}

\end{table*}

\subsection{Search for Transient mHz Variability}
\label{sec:mhz_search}

As indicated in Table~\ref{2STAB}, the individual \textit{NICER} exposures are short and fragmented. This makes it difficult to use standard Fourier analysis to characterize low-frequency variability beyond the coherent pulse signal and its harmonics. We therefore used two complementary time--frequency methods, wavelet analysis and the Hilbert--Huang Transform (HHT), to search for possible transient mHz variability in these short good time intervals (GTIs). The purpose of this analysis is exploratory: we use these methods to identify and describe candidate low-frequency features, rather than to claim a firm QPO detection.

We first applied a continuous wavelet transform (CWT) to the
\textit{NICER} light curves, following the formulation of
\citet{torrence1998practical}. For a discrete time series $x_{n'}$,
the CWT is defined as
\begin{equation}
W_n(s)=\sum_{n'=0}^{N-1}x_{n'}
\Psi^*\left[\frac{(n'-n)\delta t}{s}\right],
\end{equation}
where $N$ is the number of time samples, $n$ and $n'$ are time
indices, $\delta t$ is the sampling interval, $s$ is the wavelet
scale, $\Psi$ is the wavelet basis function, and the asterisk denotes
the complex conjugate.

We adopted the complex Morlet wavelet as the mother wavelet,
\begin{equation}
\Psi_0(\eta)=\pi^{-1/4}
e^{i\omega_0\eta}e^{-\eta^2/2},
\end{equation}
where $\eta$ is a dimensionless time parameter and $\omega_0$ is the
dimensionless central frequency. The Morlet wavelet provides a useful
balance between time and frequency resolution for oscillatory
signals. We adopted the standard value $\omega_0=6$, for which the
small non-zero-mean correction is negligible and the wavelet scale
is close to the corresponding Fourier period
\citep{torrence1998practical,zhu2026timing}.

Following the normalization convention of
\citet{torrence1998practical}, the Fourier transform of the scaled
wavelet is written as
\begin{equation}
\hat{\Psi}(s\omega_k)
=
\left(\frac{2\pi s}{\delta t}\right)^{1/2}
\hat{\Psi}_0(s\omega_k),
\end{equation}
such that
\begin{equation}
\sum_{k=0}^{N-1}
\left|\hat{\Psi}(s\omega_k)\right|^2=N.
\end{equation}
Here, the hat denotes a Fourier transform and $\omega_k$ is the
angular Fourier frequency. Using the convolution theorem, the CWT
can then be evaluated in Fourier space as
\begin{equation}
W_n(s)=\sum_{k=0}^{N-1}
\hat{x}_k\hat{\Psi}^*(s\omega_k)
e^{i\omega_k n\delta t},
\end{equation}
where $\hat{x}_k$ is the discrete Fourier transform of the time
series. The localized wavelet power is given by
$P_n(s)=|W_n(s)|^2$. Further details and applications of wavelet
analysis in X-ray timing can be found in \citet{zhu2025timing}.

\begin{figure*}
\centering
    \includegraphics[width=0.48\columnwidth]{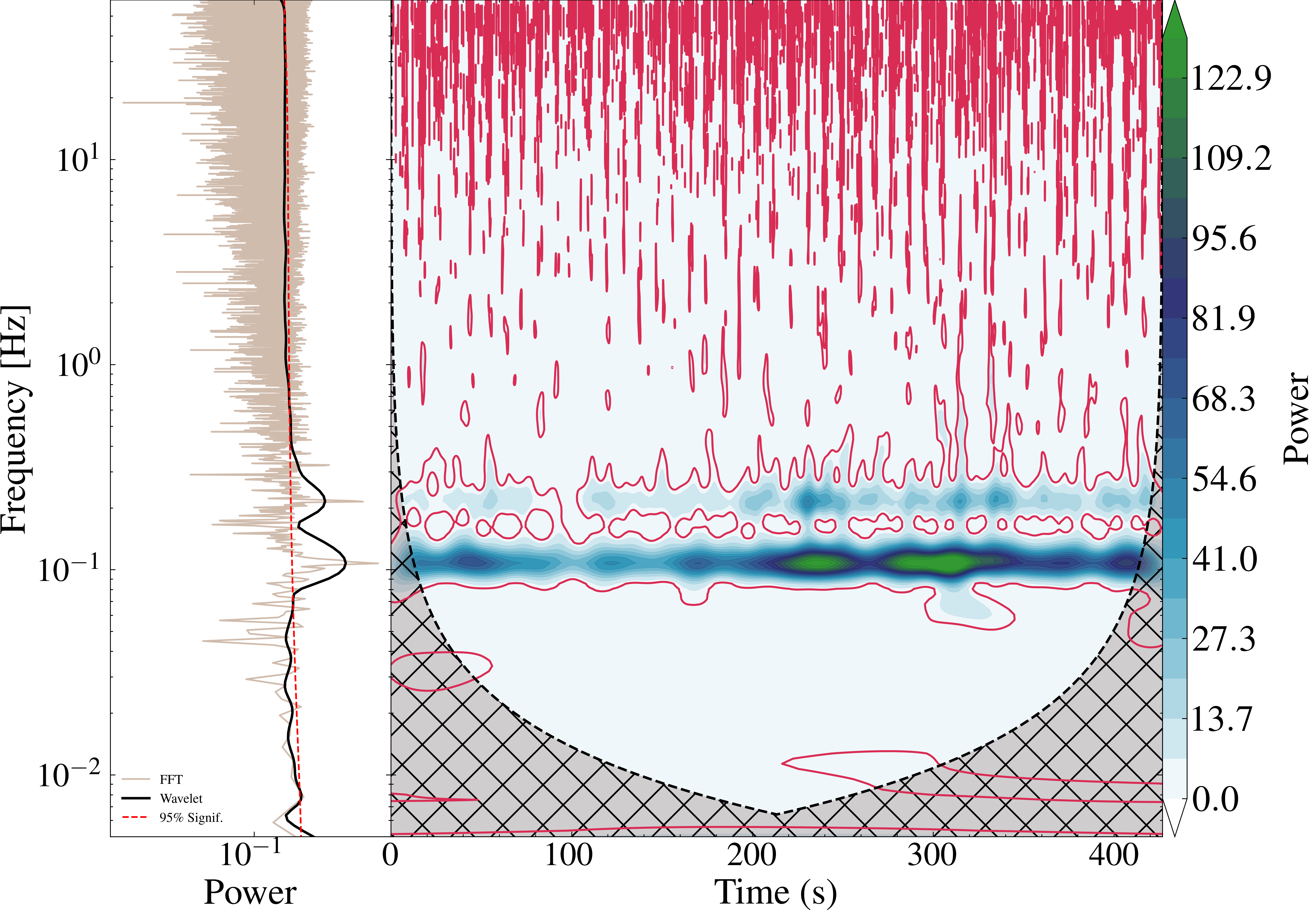}
    \includegraphics[width=0.48\textwidth]{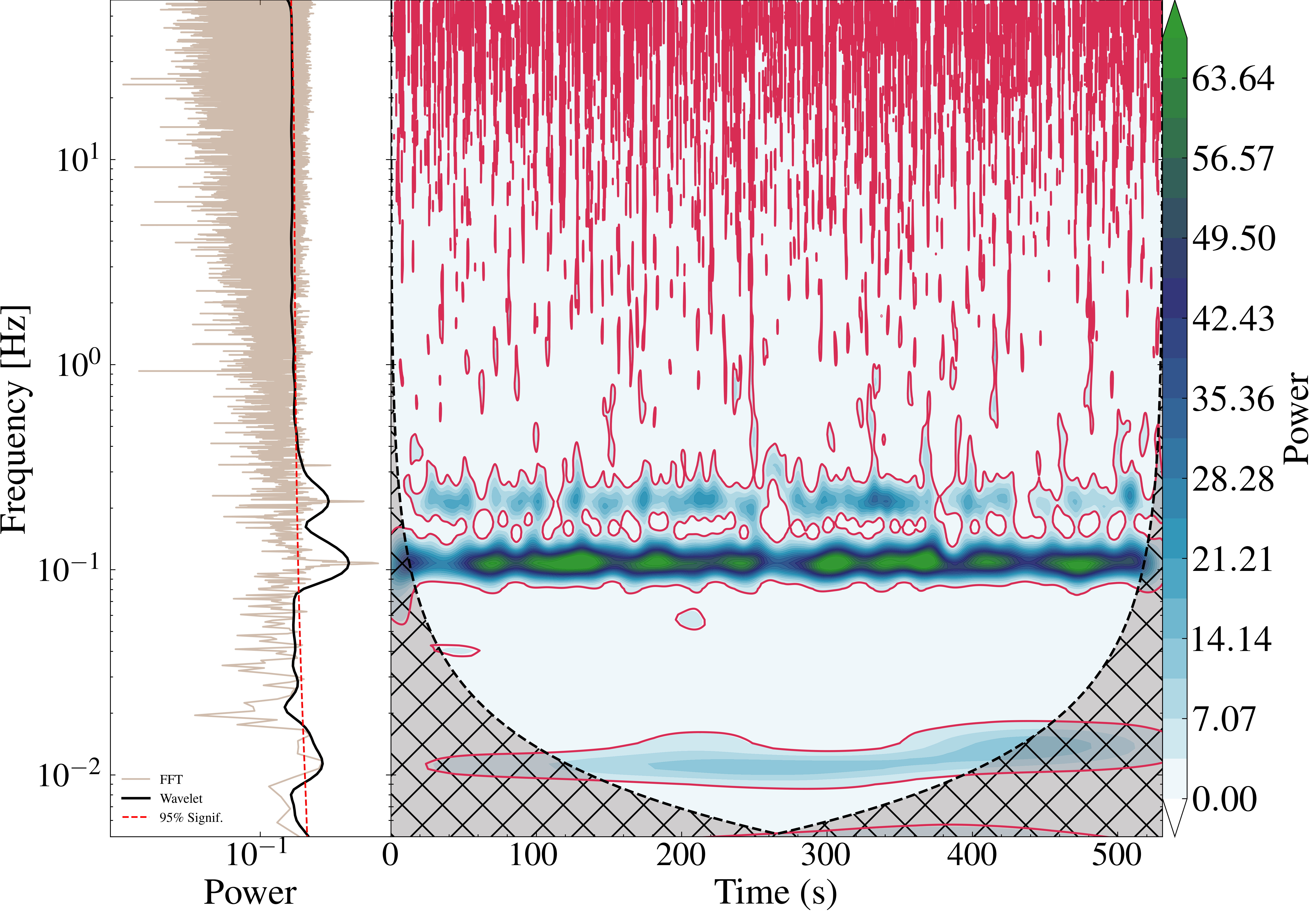}\\
    \includegraphics[width=0.48\textwidth]{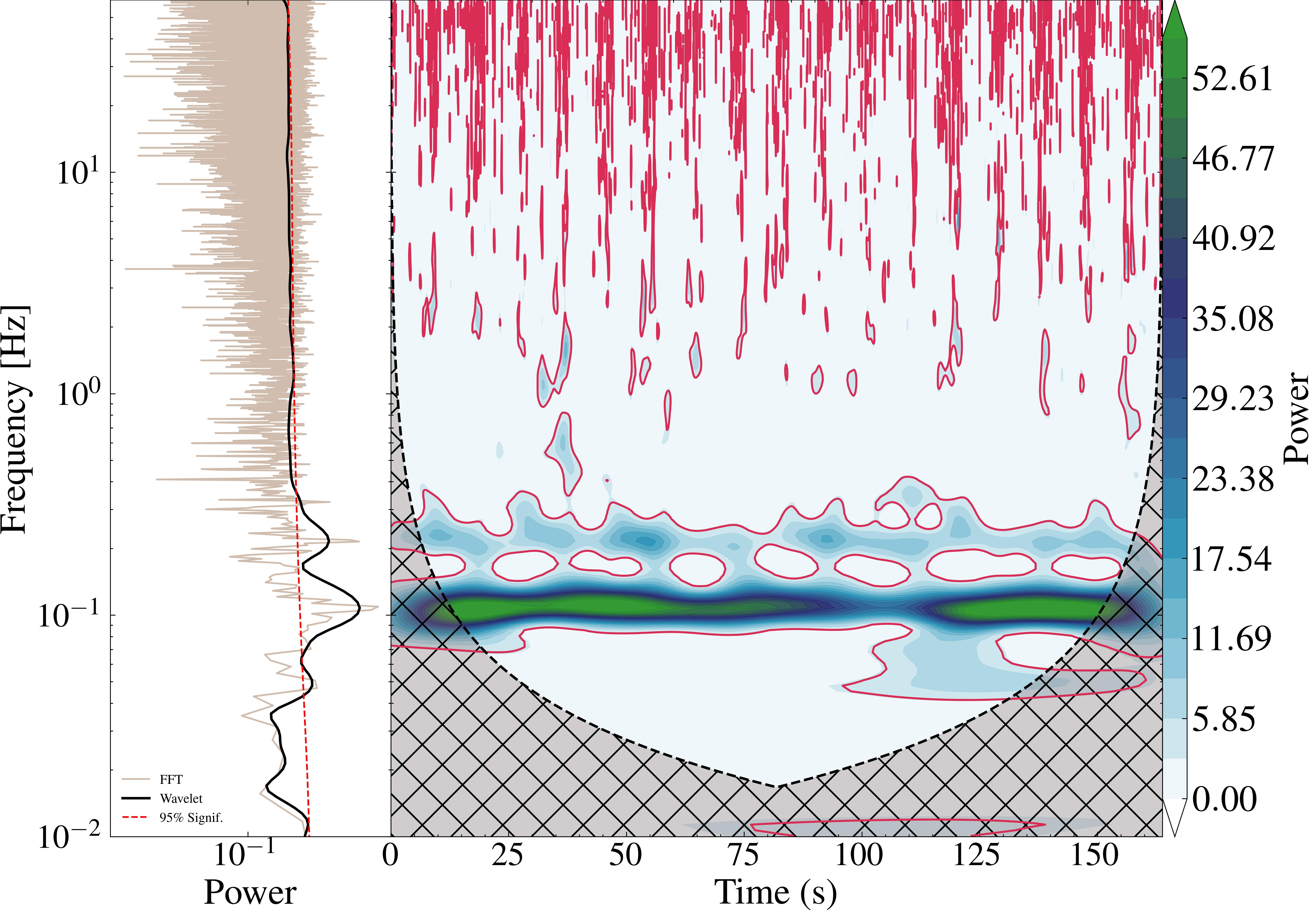}
    \includegraphics[width=0.48\textwidth]{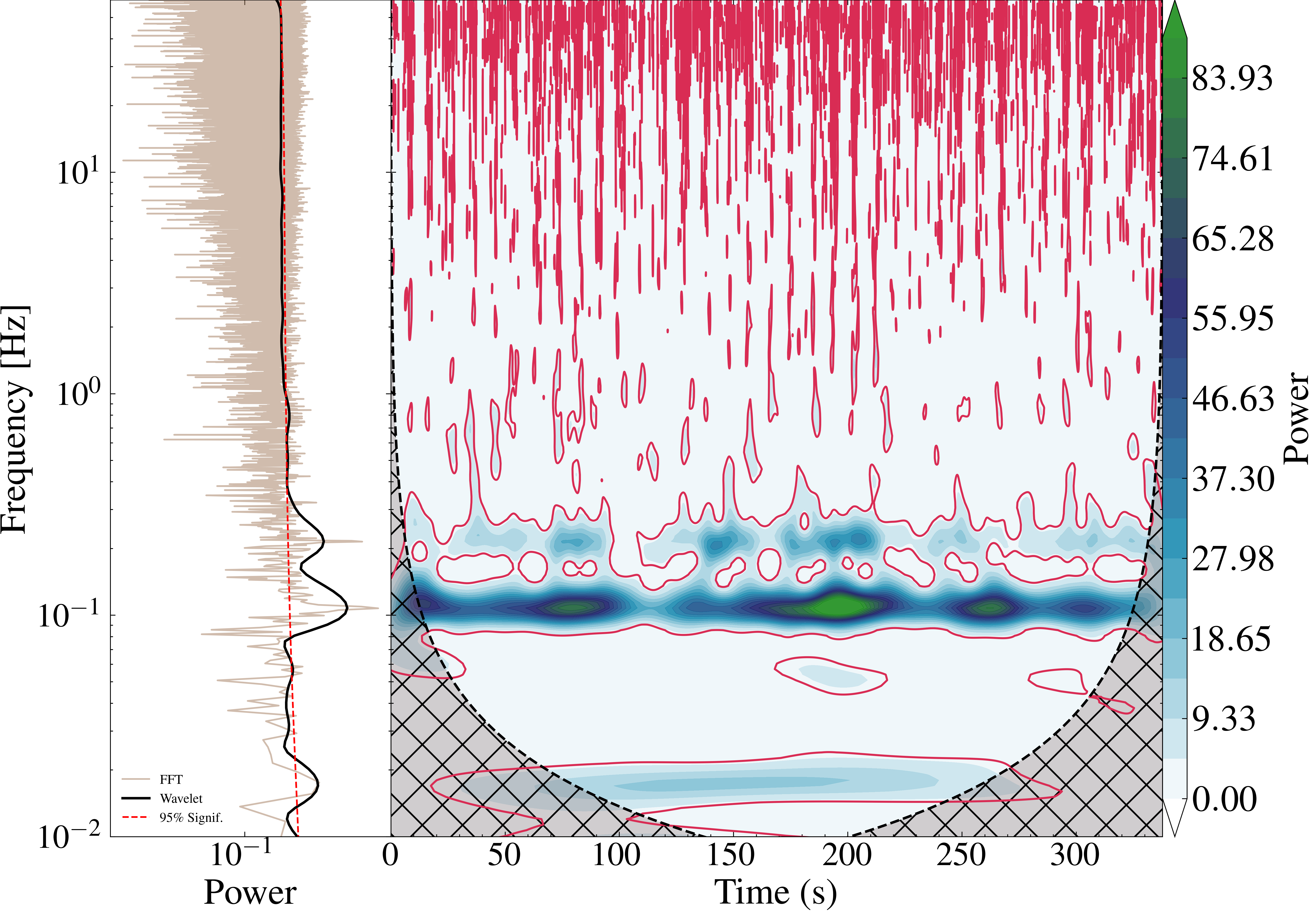}
    \caption{Wavelet power spectra for several representative \textit{NICER} GTIs. Upper-left: the third GTI of ObsID 7202030101. Upper-right: the fourth GTI of ObsID 7202030101. Lower-left: the first GTI of ObsID 7202030103. Lower-right: the first GTI of ObsID 7202030104. Localized low-frequency excesses are seen in the two GTIs shown in the right column. The color scale indicates the wavelet power, with brighter colors corresponding to stronger variability. The shaded gray region marks the COI, where edge effects make the wavelet power less reliable. The vertical axis shows the Fourier-equivalent frequency, the left horizontal axis indicates the wavelet power, and the bottom horizontal axis denotes time. The red contours represent the 95\% confidence level. All wavelet transforms were computed using the Morlet wavelet with $\omega_{0}=6$.}
    \label{wave}
\end{figure*}

The analysis was carried out with the \texttt{pycwt}\footnote{https://github.com/regeirk/pycwt} package. We processed all \textit{NICER} GTIs with durations longer than 100~s, using a time resolution of $0.0078125$~s. Representative wavelet power spectra are shown in Fig.~\ref{wave}. Most GTIs show only the coherent pulse frequency and its harmonics. In two GTIs, however, we find localized low-frequency excesses near $\sim10$~mHz and $\sim20$~mHz. These excesses are intermittent and only marginally exceed the adopted confidence level. In addition, a substantial fraction of the low-frequency power lies close to, or within, the cone of influence (COI), where boundary effects can affect the reliability of the wavelet power. We therefore treat these features as candidate transient mHz variability, rather than as robust QPO detections.

To further examine these candidate features, we also applied the HHT. The HHT was used here as an adaptive decomposition tool to inspect whether oscillatory components appear in the same frequency ranges suggested by the wavelet maps. It should not be interpreted as an independent formal detection test. The HHT analysis consists of two steps. First, the light curve $x(t)$ is decomposed into a set of intrinsic mode functions (IMFs) and a residual trend $r_n(t)$,
\begin{equation}
    x(t) = \sum_{j=1}^{n} c_j(t) + r_n(t),
    \label{eq:emd}
\end{equation}
where $c_j(t)$ is the $j$-th IMF component. We used the Complete Ensemble Empirical Mode Decomposition with Adaptive Noise (CEEMDAN) algorithm to reduce mode mixing \citep{torres2011complete}. In our implementation, we performed 1000 ensemble trials and added white noise with an amplitude of $0.2$ times the standard deviation of the original light curve. To reduce end effects, a mirror extension was applied to both ends of the light curve before the decomposition \citep{rilling2003empirical,zeng2004simple}.

Second, we applied the Hilbert transform, $\mathcal{H}[\cdot]$, to each IMF to construct the analytic signal,
\begin{equation}
    z_j(t) = c_j(t) + i \mathcal{H}[c_j(t)] = a_j(t) e^{i \phi_j(t)},
    \label{eq:analytic_signal}
\end{equation}
where $a_j(t)$ is the instantaneous amplitude and $\phi_j(t)$ is the instantaneous phase. The instantaneous frequency is then defined as $\omega_j(t)=d\phi_j(t)/dt$. This representation allows us to follow the characteristic frequency and amplitude of each IMF as a function of time \citep{huang1998empirical,zhu2025timing}.

For the low-frequency search, we extracted barycenter-corrected photon events in the 1--10~keV energy band from the GTIs where the wavelet maps showed candidate mHz excesses. We then constructed light curves with a time resolution of $\Delta t=1$~s. This time resolution is adequate for variability in the $\sim0.001$--$0.02$~Hz range, corresponding to timescales of tens to hundreds of seconds. From the CEEMDAN decomposition, we inspected the IMFs whose characteristic frequencies fall in the frequency interval indicated by the wavelet analysis. The instantaneous amplitude $A(t)$ and instantaneous frequency $\nu(t)$ were then derived from the selected IMF.

\begin{figure}
\centering
    \includegraphics[width=0.8\columnwidth]{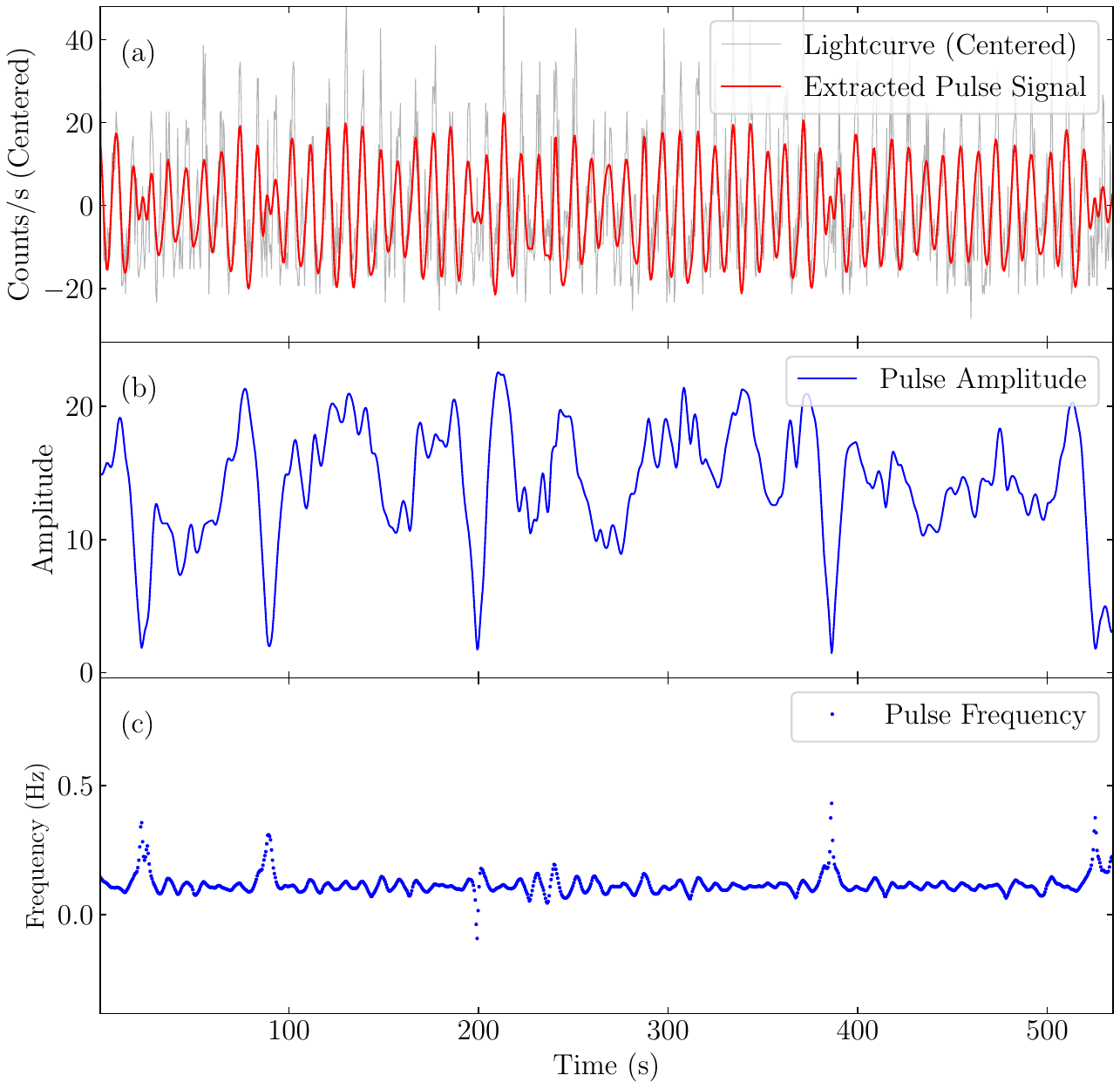}
    \caption{Pulse signal extraction and time-frequency analysis based on the CEEMDAN--HHT method for the fourth GTI of ObsID 7202030101. (a) Comparison between the original mean-centered light curve (gray thin line) and the extracted pulse component (red solid line). (b) Time evolution of the instantaneous amplitude of the pulse component. (c) Time evolution of the instantaneous frequency.}
    \label{hhtpulse}
\end{figure}

As a check on the decomposition, we first applied the same procedure to the coherent pulse signal in the fourth GTI of ObsID 7202030101. For this test, the time resolution was set to $0.5$~s to better sample the pulse period. The results are shown in Fig.~\ref{hhtpulse}. The IMF with a characteristic frequency around the pulse frequency successfully follows the coherent pulsation, demonstrating that the adopted decomposition can recover a known periodic component from the short light curve.

\begin{figure}
\centering
    \includegraphics[width=0.8\columnwidth]{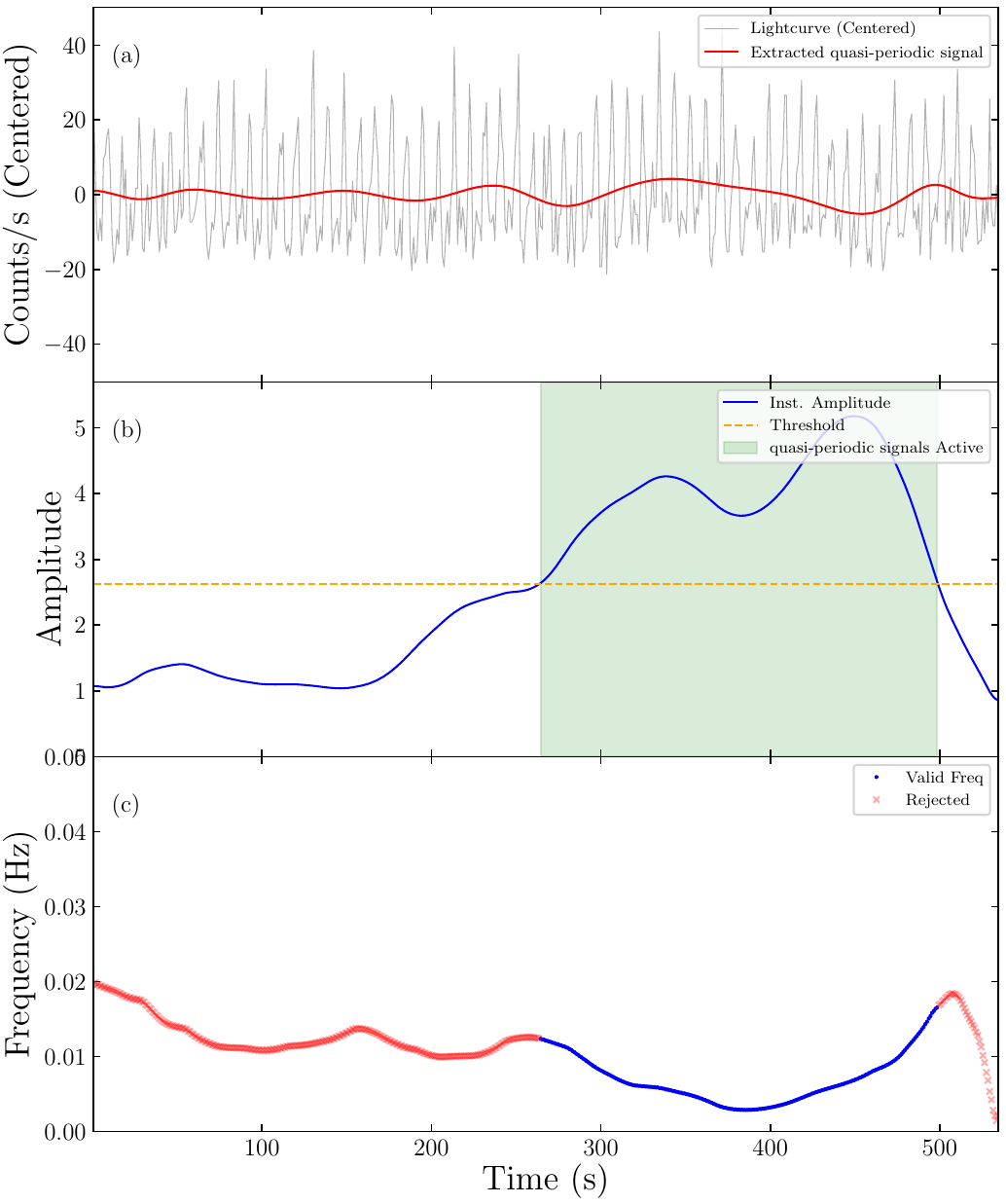}
    \caption{CEEMDAN--HHT analysis of the candidate mHz variability in the fourth GTI of ObsID 7202030101. (a) The mean-centered light curve with a time resolution of 1~s, overlaid with the selected low-frequency IMF. (b) Instantaneous amplitude of the selected IMF. The horizontal dashed line marks the empirical threshold $A(t)=\langle A\rangle$, used only to visualize intervals where the extracted component is relatively strong. (c) Instantaneous frequency evolution. Blue points show intervals above the amplitude threshold, while red crosses indicate intervals below the threshold.}
    \label{mhz0}
\end{figure}

\begin{figure}
\centering
    \includegraphics[width=0.8\columnwidth]{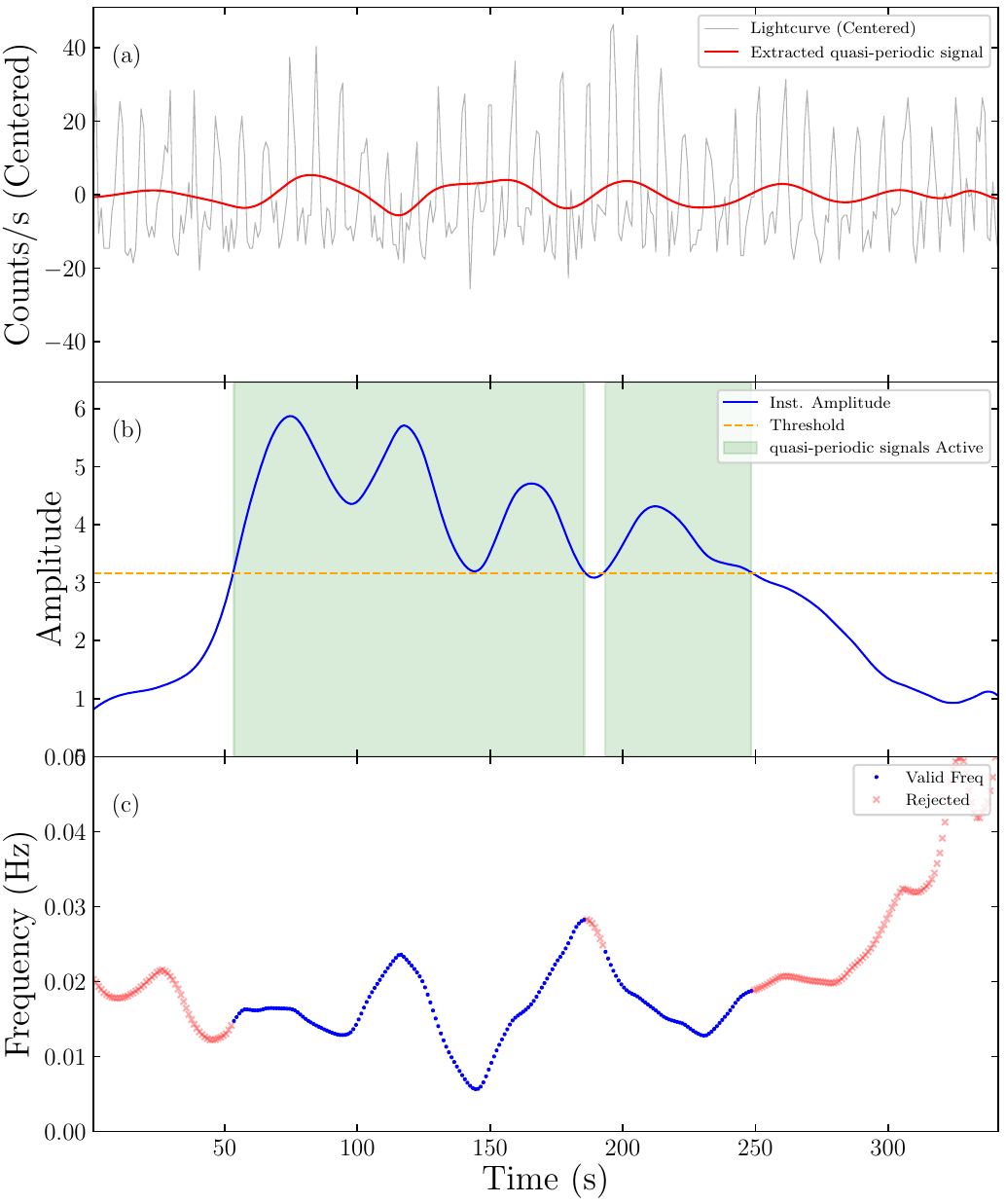}
    \caption{Same as Fig.~\ref{mhz0}, but for the first GTI of ObsID 7202030104. The selected low-frequency component corresponds to IMF~4.}
    \label{mhz1}
\end{figure}

We then applied the same procedure to the candidate mHz features. The results are shown in Figs.~\ref{mhz0} and \ref{mhz1}. The details of the decomposed components are presented in Figs.~\ref{imf0} and \ref{imf1}. For the fourth GTI of ObsID 7202030101, the selected IMF has a characteristic frequency around $\sim10$~mHz. For the first GTI of ObsID 7202030104, the selected IMF lies near $\sim20$~mHz. These frequencies are consistent with the localized excesses seen in the wavelet power spectra. We used the instantaneous amplitude only to identify intervals where the selected IMF is relatively strong, adopting the empirical criterion $A(t)>\langle A\rangle$. This threshold is used as a practical way to visualize the intermittent behavior of the extracted component; it is not a formal statistical significance criterion.

\begin{figure}
    \includegraphics[width=0.7\columnwidth]{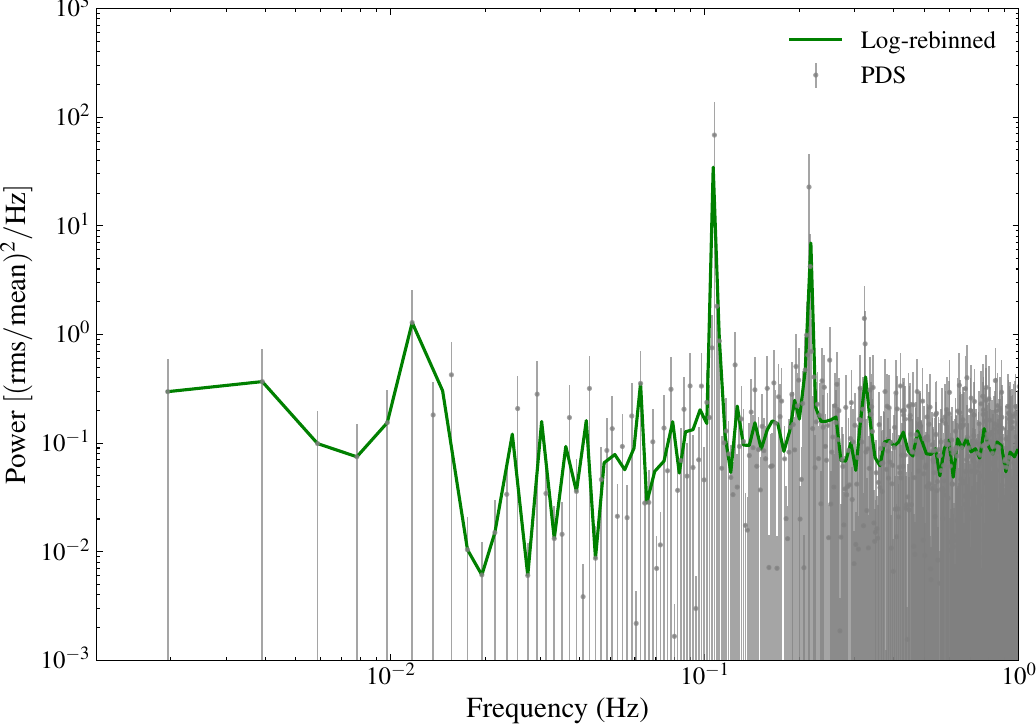}
    \caption{FFT-based power density spectrum (PDS) derived from the
fourth GTI of ObsID 7202030101. The gray points with error bars show
the raw PDS, while the green line shows the logarithmically rebinned
PDS. A segment length of 512~s was adopted to improve the frequency
resolution around 10~mHz. Because only one short segment is
available, the uncertainties remain large, and the candidate
low-frequency feature cannot be meaningfully constrained.}
    \label{PDS}
\end{figure}

For comparison with standard Fourier analysis, Fig.~\ref{PDS} shows the FFT-based power density spectrum (PDS) derived from the fourth GTI of ObsID 7202030101. A segment length of 512~s was adopted to provide frequency resolution in the mHz range. However, the short continuous exposure allows only a single segment to be used. The candidate $\sim10$~mHz variability corresponds to a timescale of $\sim100$~s, so the GTI contains only a few cycles. As a result, the PDS has large uncertainties and the low-frequency feature cannot be meaningfully constrained. Fitting the PDS with a constant, a red-noise component, and Lorentzian components for the pulse and candidate mHz variability gives a poorly constrained centroid frequency of $9.56 \pm 17.31$~mHz for the low-frequency component. The Fourier PDS therefore does not provide a robust detection or reliable parameter constraints for the candidate feature.

Overall, the wavelet and HHT analyses provide a consistent exploratory description of localized low-frequency variability in two short \textit{NICER} GTIs. The wavelet transform identifies the approximate time--frequency regions where the excess power appears, while the CEEMDAN--HHT decomposition recovers oscillatory components in similar frequency ranges. However, because of the short exposure, the influence of the COI, red-noise fluctuations, and the lack of a well-constrained Fourier peak, we regard the $\sim10$~mHz and $\sim20$~mHz features as candidate mHz variability only. 
\section{DISCUSSION}
\label{DISCUSSION}
\subsection{Timing properties}
The energy dependence of pulse profiles provides key insights into the accretion geometry. Previous studies of 2S 1553-542 during the 2015 and 2021 outbursts identified an  X-ray wing structure near the cyclotron line energy, attributed to phase lags caused by the superposition of pencil- and fan-beam emission patterns \citep{tsygankov2016nustar,malacaria2022accreting,ferrigno20114u}. In our 2024 observations, the wing-like structure shows a pronounced
energy dependence: it is not clearly visible at 3--5~keV, becomes
weakly discernible at 5--8~keV, is most prominent at 12--22~keV,
weakens at 22--34~keV, and is no longer evident above 34~keV.  Furthermore, \textit{NICER} data indicated the structure appears only after October 7, suggesting a luminosity-dependent evolution similar to the high-luminosity behavior reported by \citet{malacaria2022accreting}, where the hard X-ray wing persists across broader energy bands compared to lower-luminosity states. This implies that the source was likely in a transition or supercritical regime where the radiation-dominated shock modifies the beaming pattern \citep{basko1976limiting}.

The PF remains high ($>60\%$) and increases monotonically with
energy (Fig.~\ref{pf}). This behavior differs from the pronounced PF
dip around the cyclotron-line energy observed during the
lower-luminosity 2015 outburst of 2S~1553$-$542
\citep{tsygankov2016nustar}, as well as in sources such as
1A~0535+262 and RX~J0440.9+4431
\citep{wang2022timing,li2023timing}. Such variations can be related
to luminosity-dependent changes in the emission geometry and
resonant scattering around the cyclotron energy.

Adopting a distance of 16--24~kpc, the measured flux corresponds to
$L_{\rm X}\approx(3.3--7.3)\times10^{37}~{\rm erg~s^{-1}}$,
which is comparable to or higher than the  critical
luminosity of
$L_{\rm crit}\approx 4.8\times10^{37}~{\rm erg~s^{-1}}$ estimated
from the CRSF-derived magnetic field. The source was therefore
likely close to the transition between the two accretion regimes
and could have occupied the high-luminosity branch, particularly
toward the upper end of the distance range. The absence of a PF dip,
together with the observed pulse-profile morphology, suggests that
the source was in a different accretion state from that observed
during the 2015 outburst.
\subsection{Spectral Properties}

The phase-averaged CRSF is well described by both \texttt{gabs}
and \texttt{cyclabs}, which give
$E_{\rm gabs}=27.95\pm0.34$~keV and
$E_{\rm cyc}=24.11\pm0.23$~keV, respectively. These values agree
with the measurements obtained during the 2015 outburst
\citep{tsygankov2016nustar} and the \texttt{gabs} centroid measured
in 2021 \citep{malacaria2022accreting,rai2023spectral}. Assuming
$z=0.3$, the \texttt{cyclabs} centroid implies a magnetic-field
strength of approximately $3\times10^{12}$~G. We therefore find no
evidence for a large long-term change in the magnetic field of the
line-forming region, although the inferred values depend on the
adopted continuum and line profile.

The continuum varies appreciably with pulse phase. The blackbody
temperature changes from approximately 0.76 to 0.95~keV, reaching
its maximum during the wing phase (0.4--0.6) and its minimum during
phase 0.8--1.0. The blackbody normalization follows the opposite
trend and is largest in the final phase bin. During the wing phase,
the photon index is also most negative and the cutoff energy reaches
its minimum. The \texttt{cutoffpl} normalization varies by a factor
of several and is highest during phase 0.8--1.0. These changes show
that the continuum shape evolves over the pulse cycle, but the
covariance among the continuum parameters makes it difficult to
assign a unique physical origin to each variation.

The CRSF is constrained in four phase bins. Its centroid energy,
width, and strength vary over 25.2--29.3~keV, 4.4--6.6~keV, and
3.7--13.1~keV, respectively. The centroid is higher in Phases~1
and 4 and lower in Phases~2 and 5, with no simple monotonic relation
to the pulse intensity. During the wing phase, the CRSF parameters
cannot be constrained, and the spectrum is fitted without the
\texttt{gabs} component. We do not interpret this result as evidence
that the line is physically absent. Its contrast may instead be
reduced by the viewing geometry or by degeneracy with the continuum.

Previous studies also found phase-dependent CRSF parameters in
2S~1553$-$542, although the detailed patterns differ among the
2015, 2021, and 2024 outbursts
\citep{tsygankov2016nustar,rai2023spectral}. Differences in
luminosity, phase definition, spectral model, and phase resolution
may all contribute. Similar observation-dependent behavior has been
reported in other accreting pulsars
\citep{liu2022variations,yang2023discovery,
yang2024evidence,yang2025broad}, supporting an interpretation in
which the line-forming region is viewed from different angles as the
neutron star rotates.

The iron line is much less variable. Its centroid remains between
6.23 and 6.35~keV, while its width ranges from 0.36 to 0.55~keV.
The normalization is also consistent within the uncertainties, and
none of these parameters shows a clear relation to the pulse
profile. The weak modulation is consistent with fluorescence in
relatively extended material surrounding the neutron star.

\subsection{Candidate mHz Variability in Short \textit{NICER} Observations}

Time--frequency methods are useful for examining low-frequency variability when the signal is intermittent or when the available exposure is short. We applied wavelet analysis and a CEEMDAN-based HHT to the short \textit{NICER} GTIs of 2S~1553$-$542 as a search for possible mHz variability. This analysis is meant as a characterization of candidate low-frequency features, not as a claim of a new QPO detection.

The wavelet maps show localized excess power near $\sim10$~mHz and $\sim20$~mHz in two GTIs. These features occur below the neutron-star spin frequency and are not obviously related to the coherent pulse or its harmonics. Their interpretation, however, is limited by the short GTI durations. At mHz frequencies, the cone of influence (COI) occupies a large fraction of the time--frequency map \citep{torrence1998practical}. Part of the low-frequency power therefore lies close to, or inside, the COI, where edge effects can affect the wavelet power and its significance. The excesses also only marginally exceed the adopted confidence level. The wavelet results therefore suggest localized low-frequency variability, but do not by themselves establish a QPO.

This is a stronger limitation than in some previous wavelet studies of mHz QPOs in X-ray pulsars. For example, \citet{yang2025observations} and \citet{yang2025detection} used longer GTIs for Her~X-1 and IGR~J19294+1816, so the COI had a smaller effect on the mHz range. In our data, the short GTIs, red-noise fluctuations, and possible intermittency make the wavelet result less secure. The data are therefore better suited to identifying candidate mHz variability than to measuring QPO parameters.

We then used the CEEMDAN--HHT analysis to inspect the same features in a different way. The HHT is adaptive and does not require a fixed basis or a fixed time window, which makes it useful for signals with time-dependent amplitude and frequency \citep{huang1998empirical,hsieh2020phase}. The CEEMDAN decomposition gives low-frequency IMFs with characteristic frequencies close to those indicated by the wavelet maps. The fourth GTI of ObsID~7202030101 contains a component near $\sim10$~mHz, and the first GTI of ObsID~7202030104 contains a component near $\sim20$~mHz. This agreement suggests that the low-frequency excesses are not produced by a single isolated time bin, but correspond to oscillatory components present during part of the GTIs.

The HHT result should still be treated with caution. The instantaneous-amplitude condition $A(t)>\langle A\rangle$ is an empirical selection criterion. It is useful for showing when the extracted low-frequency IMF is relatively strong, but it is not a formal global significance test. Likewise, CEEMDAN separates the light curve into components with different characteristic timescales, but this does not prove that the low-frequency component is physically independent of the other variability. It only shows that, within this decomposition, the candidate mHz variability appears in lower-frequency IMFs, whereas the coherent pulse and its harmonics are mainly contained in higher-frequency modes.

The standard Fourier PDS illustrates the same limitation. For the fourth GTI of ObsID~7202030101, the candidate $\sim10$~mHz feature corresponds to a timescale of about $\sim100$~s, so the GTI contains only a few cycles. Only one 512~s segment can be used, and the PDS cannot be averaged over multiple segments. As a result, the uncertainties are large. A Lorentzian component near 10~mHz is poorly constrained, with a centroid frequency of $9.56\pm17.31$~mHz. Thus, the Fourier PDS does not show a well-defined peak and cannot support a secure mHz QPO detection.

The wavelet and HHT results are therefore consistent, but not conclusive. The wavelet maps identify the approximate time and frequency ranges of the excess power, and the CEEMDAN--HHT decomposition recovers components with similar characteristic frequencies. However, the short exposures, the proximity of part of the signal to the COI, red-noise fluctuations, and the lack of a well-constrained Fourier peak all limit the significance of the result. We therefore refer to the $\sim10$~mHz and $\sim20$~mHz features as candidate mHz variability, rather than secure mHz QPOs.

If these candidate features are intrinsic to the source, they may be related to variability near the disk--magnetosphere interaction region. Their characteristic frequencies, $0.01$--$0.02$~Hz, are well below the neutron-star spin frequency of $\sim0.11$~Hz, which makes a simple Keplerian or beat-frequency interpretation less straightforward \citep{van1987intensity,alpar1985gx5}. Magnetically driven warping, precession, or unstable disk--magnetosphere coupling could in principle produce variability on mHz timescales \citep{shirakawa2002precession,roy2019laxpc,yang2025observations}. The present data, however, cannot distinguish between these possibilities. Longer and more continuous observations are needed to test whether similar features recur, whether their centroid frequencies remain stable, and whether they are associated with spectral or pulse-profile changes.

\section{Conclusions}
\label{conclusion}

We have presented a timing and spectral analysis of the 2024 outburst of the Be/X-ray binary pulsar 2S~1553$-$542 using \textit{NuSTAR} and \textit{NICER} observations. The main results are summarized as follows:

\begin{enumerate}
    \item The \textit{NuSTAR} observation gives a pulse period of $9.285022\pm0.000001$~s. The energy-resolved pulse profiles are dominated by a single peak and show an energy-dependent wing-like structure, most clearly in the $12$--$22$~keV band. The pulsed fraction remains high, above 60\%, and increases with energy, suggesting that the source was observed in a relatively high-luminosity accretion state.

    \item The phase-averaged \textit{NuSTAR} spectrum is well described by an absorbed blackbody plus cutoff power-law continuum, together with an iron emission line and a cyclotron absorption feature. Using the \texttt{cyclabs} model, we obtain a cyclotron energy of $E_{\rm cyc}\simeq24.1$~keV, corresponding to a magnetic field strength of $B\sim3\times10^{12}$~G. Phase-resolved spectroscopy shows that the continuum and cyclotron-line parameters vary with pulse phase, with the cyclotron feature becoming poorly constrained or undetectable around the pulse-wing phase. This behavior is consistent with a viewing-angle-dependent accretion-column geometry.

    \item We searched the short \textit{NICER} GTIs for transient low-frequency variability using wavelet analysis and a CEEMDAN-based HHT. Localized excesses near $\sim10$~mHz and $\sim20$~mHz are found in two GTIs, and the HHT decomposition recovers low-frequency components in similar frequency ranges. However, the short exposures, the influence of the COI, red-noise fluctuations, and the lack of a well-constrained Fourier peak limit the significance of these features. We therefore treat them as candidate mHz variability rather than firm mHz QPO detections.
\end{enumerate}


\authorcontributions{
Conceptualization, H.Z. and W.W.; methodology, H.Z.; software, H.Z.; validation, H.Z., W.W., and M.M.; formal analysis, H.Z.; investigation, H.Z.; resources, W.W.; data curation, H.Z.; writing---original draft preparation, H.Z.; writing---review and editing, H.Z., W.W., W.Y., M.M., C.G., Z.X., and P.T.; visualization, H.Z.; supervision, W.W. and M.M.; project administration, W.W.; funding acquisition, H.Z. and W.W. All authors have read and agreed to the published version of the manuscript.
} 

\funding{This work was supported by the National Key Research and Development Program of China 
(grant No.~2021YFA0718503) and the National Natural Science Foundation of China 
(NSFC; grant No.~12133007). Haifan Zhu acknowledges support from the China Scholarship Council 
(CSC; grant No.~202506270166).}

\dataavailability{
The \textit{NuSTAR} and \textit{NICER} data analyzed in this work are publicly available from NASA's High Energy Astrophysics Science Archive Research Center (HEASARC). No new observational data were generated in this work.
}

\conflictsofinterest{The authors declare no conflicts of~interest.} 



\begin{adjustwidth}{-\extralength}{0cm}
\printendnotes[custom] 
\reftitle{References}




\appendixtitles{yes} 
\appendixstart
\appendix
\section[\appendixname~\thesection]{Decomposition Results}

\label{secapp}

To illustrate the CEEMDAN decomposition used in the low-frequency search, we show the results for the two \textit{NICER} GTIs discussed in the main text. These decompositions are used to identify the IMFs whose characteristic frequencies are close to the candidate mHz features found in the wavelet maps. They should be regarded as a complementary description of the variability components, rather than as an independent proof of a QPO detection.

\begin{figure*}[h]
\centering
    \includegraphics[width=0.8\columnwidth]{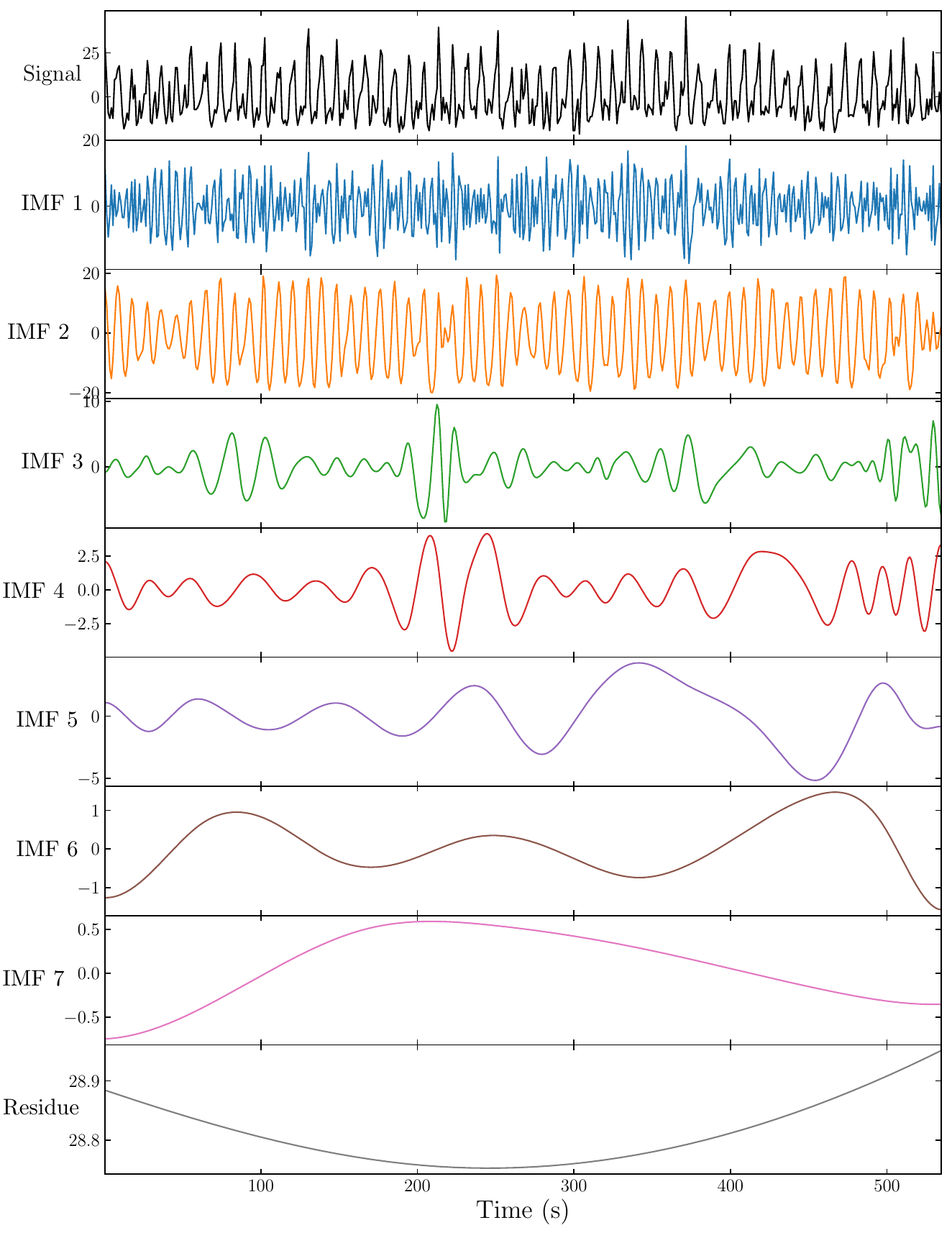}
\caption{CEEMDAN decomposition of the 1--10~keV light curve from the fourth GTI of ObsID~7202030101. The top panel shows the mean-centered light curve, followed by the extracted IMFs. The candidate $\sim10$~mHz component appears predominantly in IMF~5.}
    \label{imf0}
\end{figure*}
\begin{figure*}[h]
\centering
    \includegraphics[width=0.8\columnwidth]{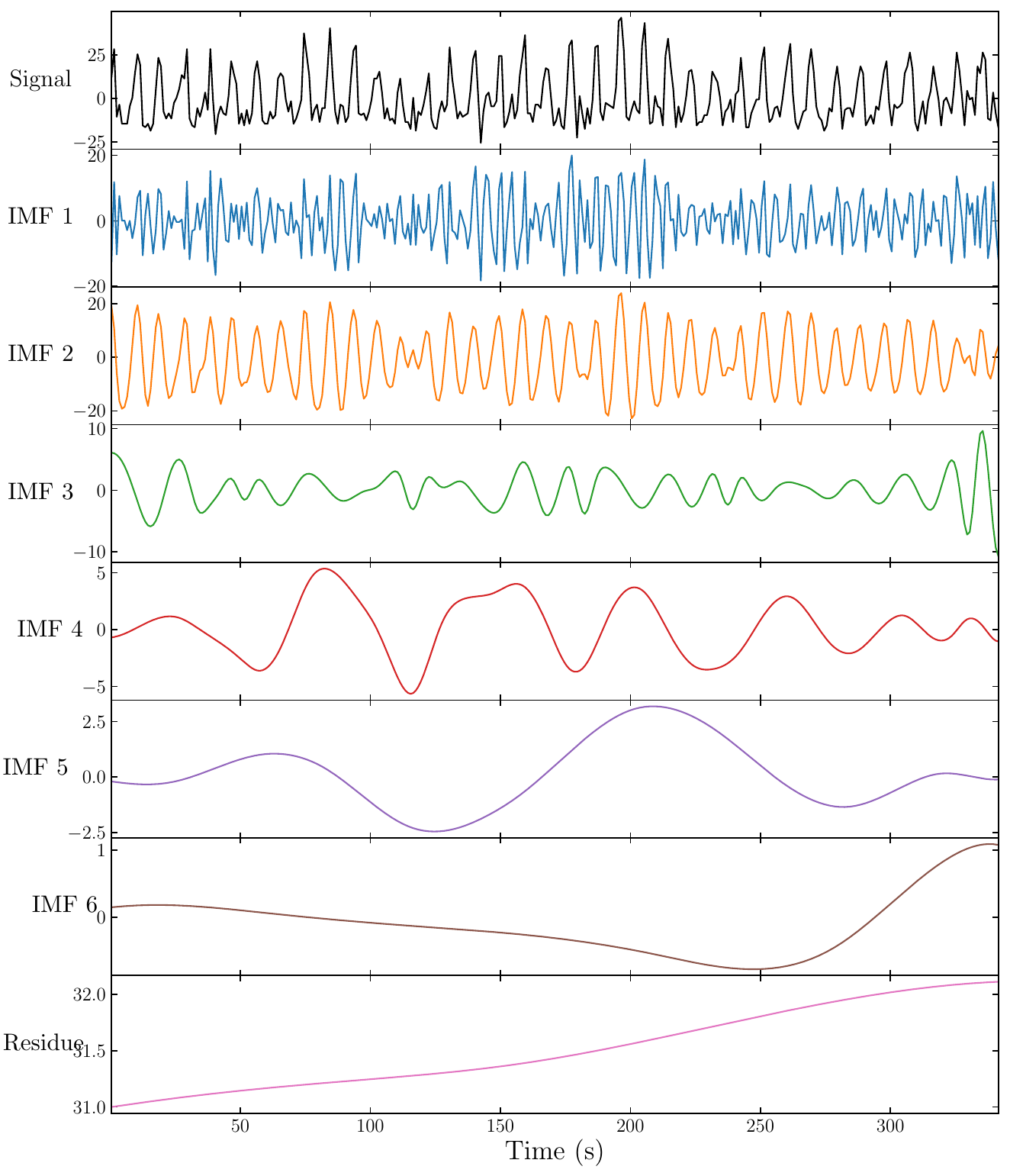}
\caption{Same as Fig.~\ref{imf0}, but for the first GTI of ObsID~7202030104. The candidate $\sim20$~mHz component appears predominantly in IMF~4. In both decompositions, the coherent pulse modulation is separated into the higher-frequency IMFs.}
    \label{imf1}
\end{figure*}
Fig.~\ref{imf0} and \ref{imf1} shows the CEEMDAN decomposition of the 1--10~keV light curves for the fourth GTI of ObsID~7202030101 and the first GTI of ObsID~7202030104. In both cases, the light curves are decomposed into IMFs with decreasing characteristic frequency, followed by a residual trend. This decomposition provides a phenomenological separation of variability on different timescales.

For ObsID~7202030101, the candidate $\sim10$~mHz variability is mainly associated with IMF~5. For ObsID~7202030104, the candidate $\sim20$~mHz variability is mainly associated with IMF~4. These components have characteristic frequencies consistent with the localized low-frequency excesses seen in the wavelet maps. The coherent pulse modulation and its harmonics are mostly contained in higher-frequency IMFs, while the candidate mHz components appear at lower frequencies.

This separation supports the view that the candidate mHz variability is not a trivial direct leakage of the coherent pulse signal. However, the decomposition alone does not establish the physical independence of the low-frequency components, nor does it provide a formal detection significance. Together with the wavelet and Fourier results discussed in the main text, the CEEMDAN decomposition is therefore used only as supporting evidence for candidate mHz variability.


\bibliography{allref}

%


\PublishersNote{}
\end{adjustwidth}
\end{document}